%% file: main.tex
\documentclass[conference]{IEEEtran}
\IEEEoverridecommandlockouts

\usepackage{cite}
\usepackage{amsmath,amssymb,amsfonts}
\usepackage{graphicx}
\usepackage{textcomp}
\usepackage{xcolor}
\usepackage{booktabs}
\usepackage{url}
\usepackage{hyperref}
\def\BibTeX{{\rm B\kern-.05em{\sc i\kern-.025em b}\kern-.08em
    T\kern-.1667em\lower.7ex\hbox{E}\kern-.125emX}}

\makeatletter
\g@addto@macro\UrlBreaks{\do\-}
\makeatother

\usepackage{algpseudocode}
\usepackage[ruled, vlined, linesnumbered]{algorithm2e}
\usepackage{graphicx}
\usepackage{textcomp}

\usepackage{listings}
\usepackage{enumitem}
\usepackage[dvipsnames]{xcolor}
\usepackage{cleveref}
\usepackage{multirow}
\usepackage{soul}
\usepackage{bm}

\usepackage{balance}

\usepackage{def}

\begin{document}
\bstctlcite{bstctl:nodash}

\title{{Enabling Spatially Fine-Grained DVFS in Neural Processing Units for Energy-Efficient LLM Serving}
\thanks{*Co-primary authors.}
}

\input{authors}

\maketitle

\input{abstract}

\begin{IEEEkeywords}
dynamic voltage and frequency scaling (DVFS), neural processing units (NPUs), large language model serving, energy efficiency.
\end{IEEEkeywords}



\input{introduction}
\input{background}
\input{design}
\input{evaluation}
\input{discussion}
\input{related_work}
\input{conclusion}
\input{ack}

\balance
\bibliographystyle{IEEEtranS}
\bibliography{ref}

\input{ae}

\end{document}

%% file: authors.tex
\author{
\IEEEauthorblockN{Yuqi Xue*}
\IEEEauthorblockA{\textit{University of Illinois Urbana-Champaign}\\
Urbana, IL, USA \\
yuqixue2@illinois.edu} \\
\IEEEauthorblockN{Corey Yu}
\IEEEauthorblockA{\textit{University of Illinois Urbana-Champaign}\\
Urbana, IL, USA \\
kaiyiyu2@illinois.edu}
\and
\IEEEauthorblockN{Jerry Wu*}
\IEEEauthorblockA{\textit{University of Illinois Urbana-Champaign}\\
Urbana, IL, USA \\
tw42@illinois.edu} \\
\IEEEauthorblockN{Jian Huang}
\IEEEauthorblockA{\textit{University of Illinois Urbana-Champaign}\\
Urbana, IL, USA \\
jianh@illinois.edu}
}





%% file: abstract.tex
\begin{abstract}


As neural processing units (NPUs) evolve rapidly to accommodate the ever-increasing compute demand of large language models (LLMs), their power consumption is becoming a limiting factor.
Our study shows that using dynamic voltage and frequency scaling (DVFS) to exploit the service-level objective (SLO) slacks is a promising way to improve NPU energy efficiency for LLM services.
As tensor operators in LLMs exhibit diverse bottlenecks across NPU components, 
it is desirable to configure the frequency separately for each component to maximize their energy efficiency.

In this paper, we develop \pname{} that enables hardware and software support for spatially fine-grained, component-level DVFS on NPUs.
\pname{} refactors the NPU core pipeline to partition components into separate V/$\bm f$ domains.
It introduces lightweight cross-domain communication mechanisms to mitigate synchronization overheads across components, and extends the NPU ISA for sub-$\bm \mu$s DVFS control.
\pname{} uses a compiler-driven two-level greedy search to co-optimize instruction scheduling and per-component V/$\bm f$ selection under SLO constraints.
We implement \pname{}'s pipeline design on an open-source NPU core to verify its functionality and evaluate the energy savings with a production-level NPU simulator with various LLMs using production traces.
\pname{} reduces energy consumption of LLM services by 16.3\%--20.5\% with 2.20\% projected area overhead on a TPUv4 chip, while preserving strict SLO guarantees.

\end{abstract}

%% file: introduction.tex
\section{Introduction}
\label{sec:intro}

Neural processing units (NPUs) are widely deployed in cloud platforms to accelerate large language model (LLM) inference today~\cite{tpuv4:isca23,aws_trainium,ascend_dvfs:asplos25}.
They have become a major contributor to power consumption in datacenters~\cite{iea2025energyai,patel2023polca,tapas:asplos25}.
Hence, it is critical for cloud providers to maximize the energy efficiency of NPUs while guaranteeing service-level objectives (SLOs) for LLM services.

\begin{figure}[t]
    \centering
    \includegraphics[width=\linewidth]{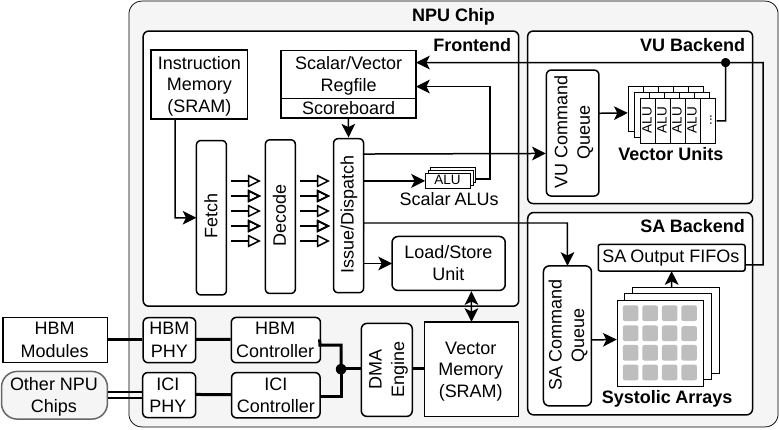}
    \caption{Architecture of an NPU chip (the connection between instruction memory and DMA engine is omitted for clarity).}
    \label{fig:npu_arch}
\end{figure}

We first conduct a study to characterize the energy consumption of LLM inference on NPUs (see \S\ref{sec:study}).
We run production LLM inference traces~\cite{dynamollm:hpca25} on NPUs with diverse LLM architectures, including both dense (Llama~\cite{llama3}) and mixture-of-experts (MoE) models (DeepSeek~\cite{deepseek}).
We quantify the energy consumption and resource utilization of all major components on an NPU 
(systolic arrays, vector units, SRAM, HBM, and ICI, as described in \Cref{fig:npu_arch} and \S\ref{sec:bkg:npu_arch}).



Our study reveals that dynamic energy accounts for 22\%--62\% of total energy, and static energy accounts for 23\%--47\% for serving various LLM models (\Cref{fig:motiv_e2e_energy_breakdown}).
The energy consumption pattern is determined by the diverse resource utilizations of different components on an NPU chip (\Cref{fig:study_overall_util}), such as the systolic arrays, vector units, SRAM, HBM, and ICI (see \S\ref{sec:bkg:npu_arch}).
Furthermore, in each LLM inference request, the utilization pattern changes over time  (\Cref{fig:study_component_util}), as different tensor operators (e.g., matrix multiplication, softmax) exhibit diverse performance demands on different components.
For example, the FFN layers during the prefill phase of a request are often systolic array-bound and underutilize vector units and SRAM bandwidth, while attention layers during the decode phase are often HBM-bound and underutilize systolic arrays.

To save static power, 
a recent study~\cite{regate:micro25} enabled power gating for NPUs. 
To save dynamic power, a well-known technique is dynamic voltage and frequency scaling (DVFS), which has been widely employed in CPUs~\cite{fine_grained_dvfs:taco11} and GPUs~\cite{dvfs_gpu:amd:asplos23} by adjusting the voltage and frequency (V/$f$) of each CPU core or GPU SM (streaming multiprocessor). 
However, DVFS for NPUs remains coarse-grained, treating the whole NPU core or chip as a single V/$f$ domain~\cite{ascend_dvfs:asplos25}. 
This inevitably misses significant energy-saving opportunities, as it cannot adjust the V/$f$ of each component in an NPU chip independently to precisely match its performance demand.
Ideally, we wish to enable \textit{spatially fine-grained DVFS}, allowing each component to operate at its energy-optimal V/$f$ state to maximize energy savings with minimal negative impact on performance. 

However, realizing spatially fine-grained (i.e., component-level) DVFS on NPUs introduces three major challenges.

First, modern NPUs do not provide architectural support for component-level V/$f$ domains.
The NPU core pipeline is tightly coupled by design, with all components coordinated through a centralized frontend (including the vector register file and a 
scoreboard for dependency tracking).
Supporting independent V/$f$ domains requires carefully refactoring the NPU core pipeline to mitigate frequency mismatches and cross-domain synchronization overhead.

Second, component-level DVFS complicates compiler scheduling. 
Compilers for NPUs statically schedule instructions, relying on deterministic instruction latencies to perform optimizations. With component-level DVFS,
instruction latencies become frequency-dependent, and different V/$f$ choices can change both instruction schedule and the bottleneck of an operator. 
As a result, the compiler must co-optimize DVFS decisions and instruction scheduling.

Third, component-level DVFS greatly expands the optimization space. Prior DVFS techniques assume a single coarse-grained compute domain and 
only decide how much to slow down each operator under a performance target. 
In contrast, component-level DVFS must jointly determine the V/$f$ state of multiple components for every operator, creating a 
search space that is 
exponentially larger.

\noindent
\textul{Our solution.}
We propose \pname{} to enable hardware and software support for component-level DVFS on NPUs.

First, we refactor the NPU core pipeline to support component-level V/$f$ domains with low hardware overhead.
\pname{} partitions the frontend, SA, VU, SRAM, HBM, and ICI into separate V/$f$ domains and introduces lightweight cross-domain communication mechanisms to preserve throughput at varying frequencies.
We carefully size asynchronous FIFOs to absorb synchronization delays and introduce a lightweight shadow vector register file to enable local forwarding within the VU domain, preventing severe pipeline stalls caused by back-to-back dependent vector instructions.

\pname{} leverages the ML compiler to manage DVFS at the tensor operator granularity. We extend the NPU ISA with a new \texttt{vf.set} command (Figure~\ref{fig:dvfs_isa}) that allows software to set the V/$f$ state of different components at sub-$\mu$s scale.
This compiler-driven approach 
exploits the highly predictable execution of ML workloads,
allowing the compiler to accurately determine bottlenecking components and estimate the performance of a scheduled VLIW code snippet.


The ML compiler must search a massive optimization space --- hundreds to thousands of operators in a request, multiple components per operator, and many V/$f$ states per component --- for a solution that does not violate SLO.
\pname{} decomposes this problem into two levels.
It first performs an operator-level search to identify near-Pareto-optimal DVFS plans that capture the near-optimal energy-delay tradeoff of each operator.
Then, it performs a request-level search to select among the Pareto plans to minimize end-to-end energy under the available SLO slack (i.e., the slowdown allowed for a request without violating its SLO)\footnote{Our study with real LLM service traces (see \S\ref{sec:study}) finds that there is abundant SLO slack, offering opportunities for us to exploit fine-grained DVFS on NPUs for energy saving without violating SLO of LLM services.}.
Since the energy-delay tradeoff of DVFS is convex ($P \propto V^2f$), \pname{} can efficiently search the Pareto plans with a simple gradient-descent search, guided by the ML compiler's instruction-level cost model that captures how DVFS changes instruction latencies and cross-domain synchronization overheads.

To integrate \pname{} into the LLM serving stack, we first perform offline analysis to identify the optimized DVFS plans for pre-compiled graphs for various batch sizes, sequence lengths, and SLO slack settings.
At runtime, the request scheduler tracks each request's SLO slack and selects the appropriate DVFS plan.
\pname{} employs simple safeguards to prevent head-of-line blocking due to V/$f$ throttling and temporarily locks NPU chips to the peak V/$f$ state to prevent SLO violations during load spikes.

We implement \pname{}'s hardware design with the open-source Coral NPU core~\cite{coralnpu}. \pname{} incurs 6.16\% area overhead on Coral NPU, and the projected area overhead is 2.20\%/2.31\% on TPUv4/TPUv5p (see \S\ref{sec:impl}).
To evaluate \pname{} with LLM workloads, we use a production-level NPU simulator that is validated against real TPU chips.
We run LLM services with various dense (Llama~\cite{llama3}) and MoE (DeepSeek~\cite{deepseek}) models using a production trace~\cite{dynamollm:hpca25}.
At a 0\% SLO slack, \pname{} saves 15.1\%/20.1\% energy for prefill/decode.
At a 10\% SLO slack, the savings increase to 20.9\%/20.1\%. Compared to the state-of-the-art NPU DVFS solution, \pname{} improves the energy savings by up to 12.6\%/7.2\% for prefill/decode.
Combining \pname{} with power gating~\cite{regate:micro25} further achieves up to 32.4\% energy savings.
We summarize our contributions as follows:
\begin{itemize}[leftmargin=*]
    \item We quantify the DVFS opportunities of LLM serving on NPUs, showing that SLO slacks and component-level underutilization create substantial room for energy savings.

    \item We propose \pname{}, a hardware-software co-design that enables spatially fine-grained, component-level DVFS on NPUs through lightweight hardware and ISA extensions.

    \item We develop a compiler-driven DVFS policy that co-optimizes instruction scheduling and per-component V/$f$ settings using an efficient two-level search algorithm.

    \item We implement and evaluate \pname{} with hardware synthesis and a production-level NPU simulator, demonstrating significant energy savings while preserving strict SLO guarantees.
\end{itemize}

%% file: background.tex
\section{Background and Motivation}
\label{sec:bkg}

\subsection{NPU Architecture}
\label{sec:bkg:npu_arch}

\noindent
\textbf{NPU core pipeline.}
Neural processing units (NPUs) are specialized machine learning (ML) accelerators.
A typical example is Google TPU~\cite{tpuv4:isca23,tpudesign:google:ieeemicro21,coralnpu}.
\Cref{fig:npu_arch} shows the NPU core pipeline.
The frontend fetches VLIW instructions and dispatches them in-order. Each VLIW instruction consists of multiple \textit{commands}.
To prevent data and structural hazards, the architecture employs a lightweight scoreboard using a wait bit per register to stall conflicting commands.
Once issued, commands are routed to specialized execution backends: systolic arrays (SAs) handle matrix multiplications and convolutions, while vector units (VUs) execute generic vector computations such as element-wise and activation functions.
Commands can retire out-of-order, overlapping execution across components.

The memory hierarchy employs an on-chip SRAM (vector memory) to exploit data reuse and hide the latency of the off-chip High-Bandwidth Memory (HBM).
A Direct Memory Access (DMA) engine orchestrates asynchronous data movement between the SRAM and HBM.
To scale out, multiple NPU chips can be interconnected via high-speed inter-chip interconnect (ICI) links to form an NPU pod (typically a 2D/3D torus~\cite{tpuv4:nsdi24}).
The DMA engine orchestrates remote DMA (RDMA)
to access other chips' HBM or SRAM.

\noindent
\textbf{NPU software stack.}
An ML program is defined as a computation graph of tensor operators~\cite{xla,jax,pytorch2,tvm}.
ML compilers apply optimizations such as operator fusion and tiling.
For each operator, the compiler generates low-level commands for each component (e.g., \texttt{push/pop} vectors from/to SAs, vector commands on VUs, \texttt{load/store} from/to the SRAM, and DMA commands for HBM/ICI access) and performs static instruction scheduling based on deterministic instruction latencies.
The compiler performs register renaming and employs optimizations such as loop unrolling and software pipelining to exploit instruction-level parallelism (ILP).

\subsection{Dynamic Voltage and Frequency Scaling}
Dynamic voltage and frequency scaling (DVFS) is a common technique for managing power and energy consumption.
It is enabled by three key primitives as discussed below.

\noindent
\textbf{Integrated voltage regulators (IVRs).}
IVRs convert a high input voltage to the localized voltage required by a power domain.
With recent packaging technologies, IVRs can be integrated within the same package as the NPU die, providing nanosecond-scale voltage transitioning times and high power conversion efficiency~\cite{dvfs_fivr:2016,dvfs_gpu:amd:asplos23}.


\noindent
\textbf{Frequency generation.}
To enable fast DVFS, IVRs are paired with a clock generator to adjust frequency in nanosecond-scale~\cite{dvfs_gpu:amd:asplos23}.
One approach is to use a phase-locked loop (PLL) to generate a reference clock and a digital frequency synthesizer (DFS) to quickly select derivative output frequencies.
An alternative approach is to use voltage-adaptive frequency-locked loops (FLLs/DFLLs), which can adjust output frequency by tracking the supply voltage.


\noindent
\textbf{Cross V/$\bm{f}$ domain synchronization.}
For voltage domain crossing, voltage level shifters~\cite{dual_range_vls:arxiv26} are
used. 
They incur negligible wiring delays (typically less than a clock cycle at maximum frequency).
For clock domain crossing,
control signals
rely on multi-stage flip-flops to mitigate metastability~\cite{bisync_fifo:nocs07} and recirculation MUX to ensure data coherency~\cite{Dave:EDN}. 
For high-throughput data buses, asynchronous FIFOs are employed, which consist of a dual-port SRAM to stage data and recirculation MUXes to synchronize control signals~\cite{bisync_fifo:nocs07}.

\input{study}
\subsection{Challenges of Component-Level DVFS}
\label{sec:challenges}

DVFS techniques are widely studied in CPUs/GPUs. They typically employ multiple V/$f$ domains across \textit{homogeneous} cores, along with a separate uncore domain (last-level cache, memory controller, etc.).
However, an NPU chip typically features a large core consisting of \textit{heterogeneous} components (SA, VU, etc.). 
Enabling component-level DVFS on NPU chips introduces unique challenges.


\noindent
\textbf{Monolithic NPU pipeline vs. independent V/$\bm{f}$ domains.}
Conventional NPU pipelines are tightly coupled: a centralized frontend dispatches commands to all components, and different
components share data via the register file.
Component-level DVFS requires decoupling them into separate V/$f$ domains and inserting asynchronous FIFOs and voltage level shifters between domains.
This requires every instruction to
cross V/$f$ domain boundaries twice (dispatch and retire), significantly inflating instruction latency.
This can severely degrade performance when dependent instructions are close to each other, such as elementwise or reduction operators on the VU, as the compiler cannot find enough independent work to hide the extra delay.
Thus, component-level DVFS requires careful refactoring of the NPU core pipeline beyond simply inserting domain-crossing logic.

\noindent
\textbf{Statically scheduled VLIW ISA vs. frequency-dependent instruction latencies.}
Modern NPUs use a statically scheduled VLIW ISA to reduce hardware complexity.
The ML compiler assumes deterministic instruction latencies to perform low-level optimizations such as loop unrolling, software pipelining, and VLIW instruction packing~\cite{tpuv4:isca23,tpudesign:google:ieeemicro21}.
With component-level DVFS, instruction latencies become frequency-dependent.
Changing one component's frequency can alter the optimal instruction schedule (e.g., changing software pipeline depth or loop unrolling factors).
This makes prior DVFS approaches ineffective, as they assume a fixed code sequence and only search for the best DVFS setting for that fixed schedule~\cite{ascend_dvfs:asplos25}.
Instead, the compiler must co-optimize instruction scheduling and DVFS decisions.


\noindent
\textbf{Complex DVFS plan search space.}
Prior compiler-assisted DVFS techniques for NPUs typically assume a single coarse-grained compute domain and decide how much each operator can slow down under a performance target~\cite{ascend_dvfs:asplos25}.
With component-level DVFS, we must determine the V/$f$ state of all components for every operator.
This expands the search space combinatorially.
As each candidate V/$f$ plan may change instruction latencies, scheduling decisions, and operator bottlenecks, the compiler cannot rely on simple heuristics over a fixed code sequence.
We need a systematic approach to efficiently explore the component-level DVFS search space.

%% file: study.tex
\subsection{DVFS Opportunities in NPUs}
\label{sec:study}

\begin{figure}[t]
    \centering
    \includegraphics[width=\linewidth]{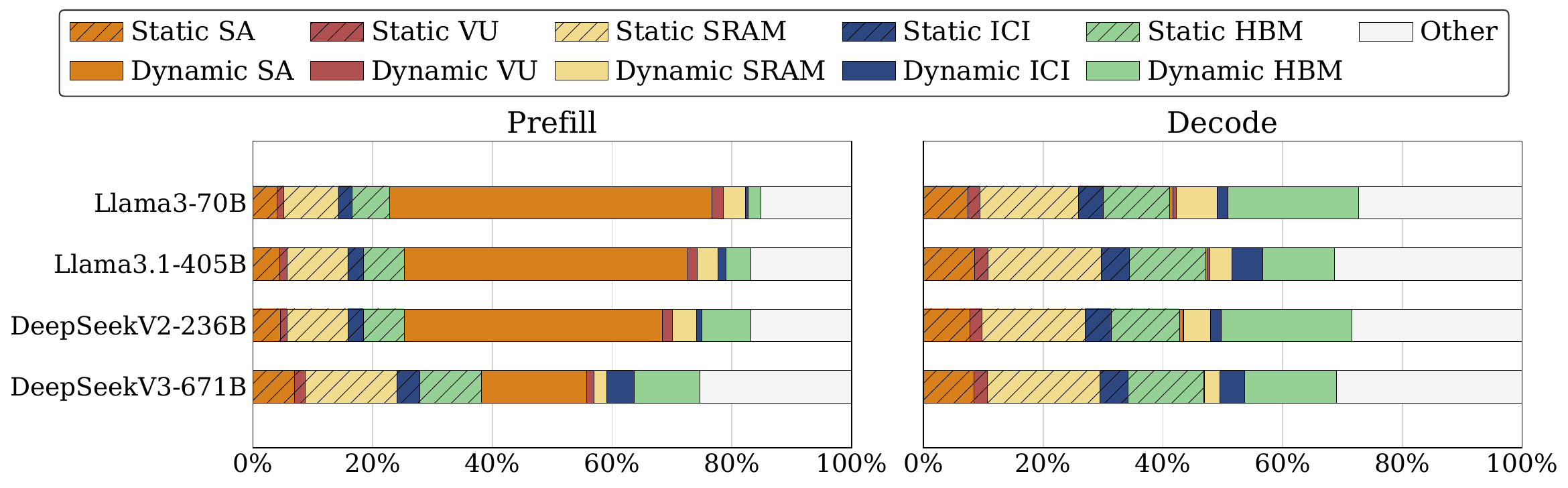}
    \caption{Energy breakdown of an LLM inference request. We use input/output sequence length 4096/512 as an example.}
    \label{fig:motiv_e2e_energy_breakdown}
\end{figure}

\begin{figure}[t]
    \centering
    \includegraphics[width=\linewidth]{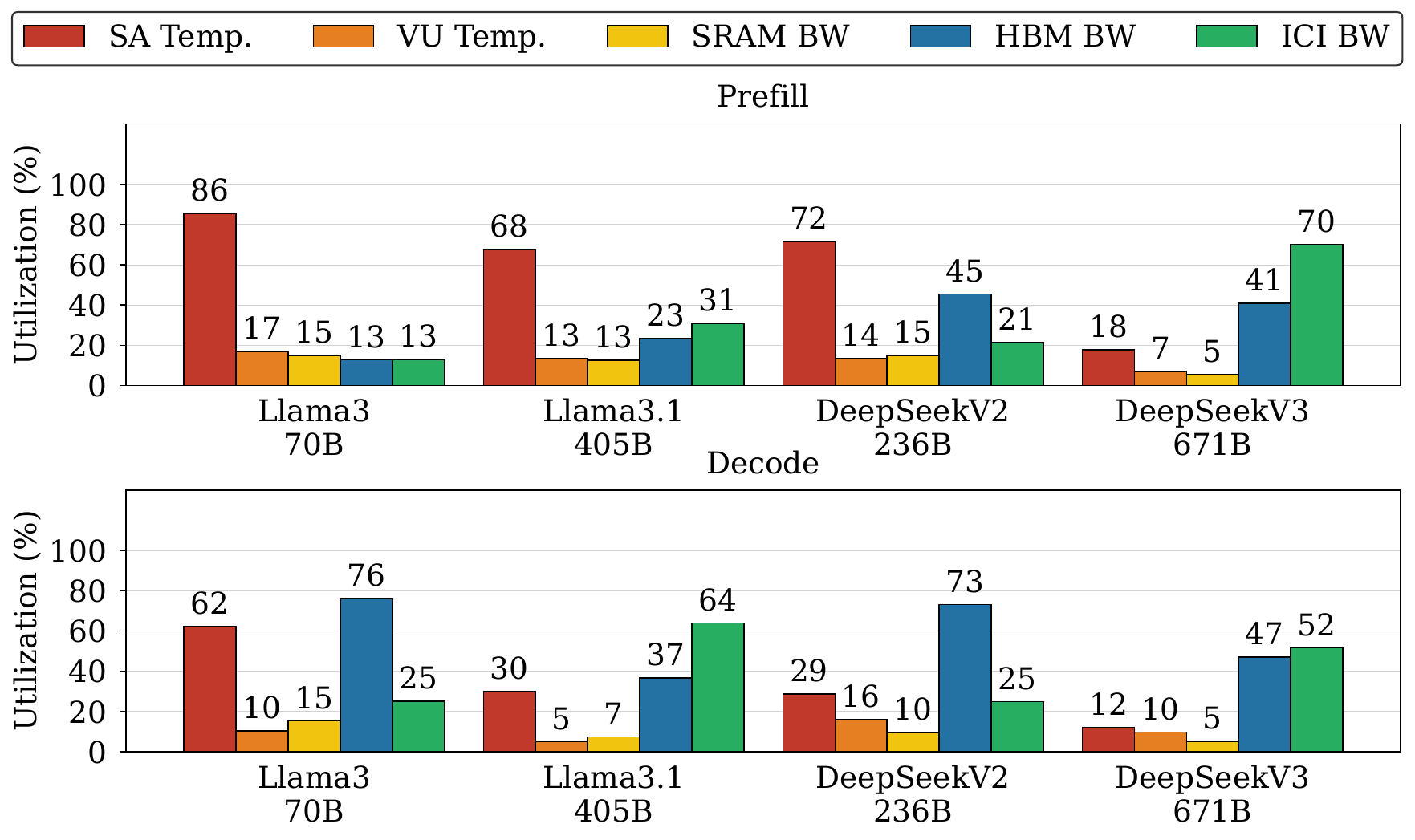}
    \caption{Per-component utilization of an LLM request. ``Temp.'': temporal utilization. ``BW'': bandwidth.}
    \label{fig:study_overall_util}
\end{figure}


In this section, we quantify the DVFS opportunities of large language model (LLM) inference services on NPUs.
We cover popular LLM architectures and different model sizes,
including dense models (Llama3~\cite{llama3}) and mixture-of-experts (MoE) models (DeepSeek~\cite{deepseek}).
As LLM prefill and decode phases have distinct resource demands, we employ state-of-the-art prefill/decode (P/D) disaggregation~\cite{DistServe:osdi24,splitwise:isca24} and study the P/D phases separately.
We use a production-level NPU simulator (see \S\ref{sec:impl}) to conduct our study.
We study the energy and resource utilization of all major components (SAs, VUs, SRAM, ICI, and HBM) when running a single LLM request. As LLM requests have diverse service-level objectives (SLOs)~\cite{dynamollm:hpca25}, we also analyze cluster-level scheduling to quantify each request's SLO slack (i.e., how much it can slow down without violating its SLO).

\begin{figure*}[t]
    \centering
    \includegraphics[width=0.95\linewidth]{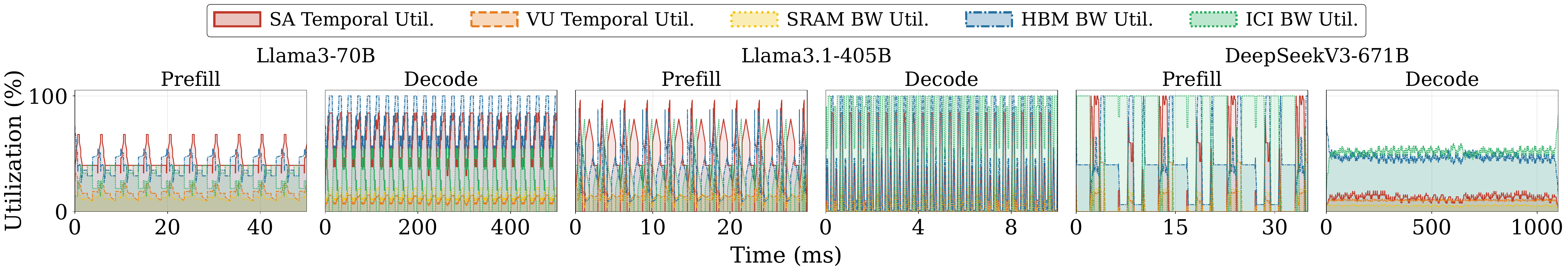}
    \caption{Utilization of each component over time. Since the transformer layers have the same architecture in each model, we show a representative time window as an example in each subplot. We do not show DeepSeekV2-236B due to space limitations.}
    \label{fig:study_component_util}
\end{figure*}

\noindent
\textbf{Energy consumption breakdown.}
We run a single request of input/output sequence length 4096/512 (see \Cref{tab:pd_node_sweep} for the P/D node configuration) and break down the component energy in \Cref{fig:motiv_e2e_energy_breakdown}.
Since prefill is compute-bound, dynamic SA energy generally accounts for over 50\% of total energy.
An exception is DeepSeekV3-671B, where MoE dispatch/combine incurs high ICI overhead.
The decode phase is memory-bound, so most dynamic energy is spent on HBM.
Static energy accounts for 23\%--47\% for prefill and decode.


\noindent
\textbf{Diverse component utilization.}
\Cref{fig:study_overall_util} shows per-component utilization, which correlates with the dynamic energy in \Cref{fig:motiv_e2e_energy_breakdown},
except that the SA can exhibit over 60\% temporal utilization but consumes little dynamic energy in decode phase.
This is because decode processes a single token at a time, incurring significant SA padding overhead and underutilizing FLOPS.
Underutilized components draw substantial static energy (23\%--38\% for prefill and 41\%--47\% for decode). Because they are often not critical to end-to-end performance, we can throttle their voltage and frequency to save energy with minimal performance degradation.


\Cref{fig:study_component_util} shows component utilization over time.
Both prefill and decode consist of diverse tensor operators, and they exhibit distinct resource demands on each component.
The average operator execution time is 5.12--291.37 $\mu$s across the evaluated models (see \Cref{tab:op_exe_time_cdf}).
This timescale makes operator-level DVFS feasible with modern IVRs that support nanosecond-scale transitions~\cite{dvfs_gpu:amd:asplos23,fine_grained_dvfs:taco11}.
We can adjust each component's V/$f$ separately to match the precise demands of each operator, saving energy with minimal performance penalty. 


\begin{table}[t]
    \centering
    \caption{1st percentile (P1), mean, standard deviation, and 99th percentile (P99) operator execution time of different LLMs. The request input/output sequence length is 4096/512.}
    \label{tab:op_exe_time_cdf}
    \begin{tabular}{@{}lp{0.8cm}ccccc@{}}
        \toprule
        \textbf{Model} & \textbf{Phase} & \textbf{P1 ($\bm{\mu}$s)} & \textbf{Mean ($\bm{\mu}$s)} & \textbf{Std ($\bm{\mu}$s)} & \textbf{P99 ($\bm{\mu}$s)} \\
        \midrule
        \multirow{2}{1.3cm}{Llama3 70B} & Prefill & 9.99 & 291.37 & 381.74 & 1079.44 \\
         & Decode & 0.50 & 9.51 & 13.03 & 39.57 \\ \midrule
        \multirow{2}{1.3cm}{Llama3.1 405B} & Prefill & 2.83 & 170.45 & 175.24 & 501.18 \\
         & Decode & 0.50 & 18.59 & 21.31 & 49.95 \\ \midrule
        \multirow{2}{1.3cm}{DeepSeekV2 236B} & Prefill & 0.50 & 10.51 & 78.95 & 173.50 \\
         & Decode & 0.50 & 5.12 & 13.59 & 98.91 \\ \midrule
        \multirow{2}{1.3cm}{DeepSeekV3 671B} & Prefill & 0.50 & 18.96 & 156.12 & 308.42 \\
         & Decode & 0.50 & 11.06 & 27.63 & 178.03 \\
        \bottomrule
    \end{tabular}
\end{table}

\noindent
\textbf{SLO slacks.}
LLM requests vary in input/output sequence lengths, which introduce diverse service-level objectives (SLOs)~\cite{dynamollm:hpca25}.
Moreover, production LLM services often over-provision resources to meet SLOs for large requests and load spikes, 
allowing performance to be traded for energy savings via DVFS without SLO violation.

To fairly quantify the exploitable SLO slack, we replay a production LLM service trace from AzurePublicDataset~\cite{dynamollm:hpca25} (see \S\ref{sec:eval:setup}). We categorize request sequence lengths into the 33rd/66th/100th percentiles and set the SLO target for each request as 5$\times$ the single-request TTFT/TPOT\footnote{Time-to-first-token (TTFT) is the prefill latency. Time-per-output-token (TPOT) is the metric for measuring decode latency.} on the minimum number of NPU chips~\cite{dynamollm:hpca25}. \Cref{tab:slo_setting} shows the SLO settings. We sweep NPU pod sizes, LLM parallelism degrees, and batch sizes to identify the most energy-efficient configuration that satisfies the SLO for all requests.
We sweep P/D node counts to determine the minimum cluster provisioning required to achieve 99\% SLO satisfaction across the trace~\cite{DistServe:osdi24,splitwise:isca24}. \Cref{tab:pd_node_sweep} shows the P/D node and NPU pod configurations.

\begin{table}[t]
    \centering
    \caption{SLO settings used in our experiments. Sequence length percentiles: P33 (input: 1{,}291; total: 1{,}313),
P66 (input: 2{,}777; total: 2{,}802),
P100 (input: 7{,}691; total: 9{,}071). For prefill, we use input lengths. For decode, we use total lengths, as the TPOT is correlated to the total KV cache size.}
    \label{tab:slo_setting}
    \begin{tabular}{l ccc ccc}
        \toprule
        \multirow{2}{*}{Model} & \multicolumn{3}{c}{TTFT (sec)} & \multicolumn{3}{c}{TPOT (millisec)} \\
        \cmidrule(lr){2-4} \cmidrule(lr){5-7}
        & P33 & P66 & P100 & P33 & P66 & P100 \\
        \midrule
        Llama3-70B      & 0.725  & 1.488  & 4.405  & 91   & 93   & 100  \\
        Llama3.1-405B   & 0.803  & 1.508  & 5.660  & 709  & 709  & 710  \\
        DeepSeekV2-236B & 0.642  & 1.302  & 5.158  & 271  & 286  & 339  \\
        DeepSeekV3-671B & 1.098  & 2.151  & 6.113  & 297  & 323  & 412  \\
        \bottomrule
    \end{tabular}
\end{table}

\begin{table}[t]
    \centering
    \caption{The prefill/decode node and cluster configurations used in our experiments. ``Config.'' is shown as data/tensor/pipeline/expert parallelism degrees.}
    \label{tab:pd_node_sweep}
    \setlength{\tabcolsep}{4pt}
    \begin{tabular}{l cc cc cc}
        \toprule
        \multirow{2}{*}{Model}
          & \multicolumn{2}{c}{Prefill Node}
          & \multicolumn{2}{c}{Decode Node}
          & \multicolumn{2}{c}{\# of Nodes} \\[-1pt]
        \cmidrule(lr){2-3} \cmidrule(lr){4-5} \cmidrule(lr){6-7}
          & Chips & Config. & Chips & Config. & P & D \\[-1pt]
        \midrule
        Llama3-70B      & 4  & 1/4/1/1  & 4  & 1/4/1/1  & 20  & 8  \\
        Llama3.1-405B   & 32 & 1/32/1/1 & 16 & 1/16/1/1 & 24  & 52  \\
        DeepSeekV2-236B & 4  & 1/4/1/1  & 4  & 1/4/1/1  & 24  & 24  \\
        DeepSeekV3-671B & 32 & 1/16/1/2 & 32 & 1/16/1/2 & 36  & 24  \\
        \bottomrule
    \end{tabular}
\end{table}

\begin{figure}[t]
    \centering
    \includegraphics[width=0.9\linewidth]{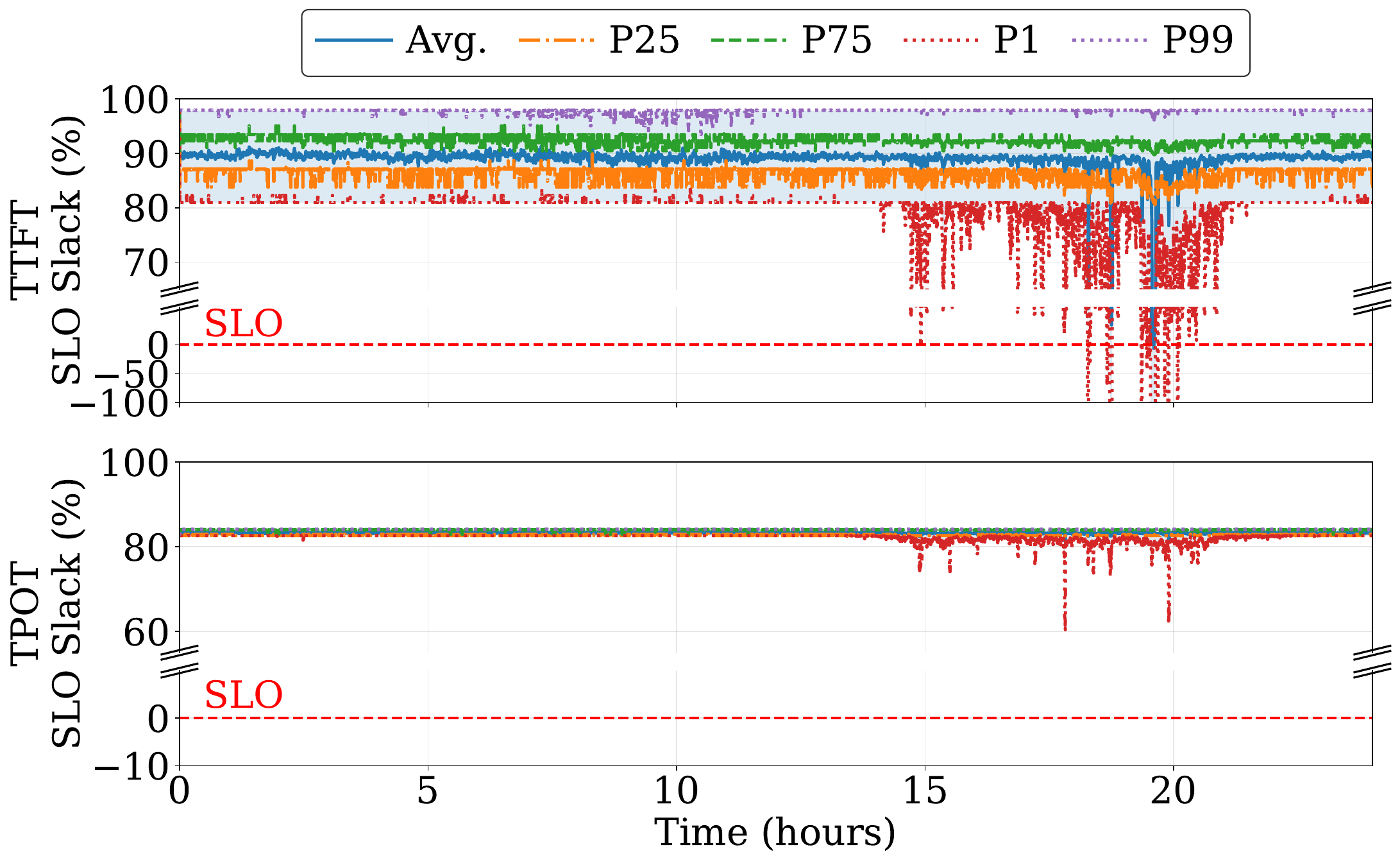}
    \caption{Normalized SLO slack time over time. For each request, the SLO slack is computed as (SLO target $-$ actual latency)$/$(SLO target). At each timestamp in the graph, the P1, P25, Avg., P75, and P99 values on the curves represent the SLO slack distribution of all requests finished in the past minute (e.g., 1-min rolling window). We plot Llama3-70B as an example (other models have a similar trend).}
    \label{fig:study_slo_slack}
\end{figure}

\Cref{fig:study_slo_slack} reports the SLO slack over time. For most requests, the TTFT/TPOT is shorter than 20\% of the SLO target. Ideally, they can be slowed down by up to 5$\times$ (for both prefill and decode) without violating SLOs.
When the request rate increases (i.e., between 14 and 22 hours), the SLO slack drops as the queuing delay increases. However, at least 75\% of requests still have more than 80\% SLO slack during both prefill and decode.
A few requests (less than 1\%) violate their SLOs and have negative slacks. This result aligns with our target of 99\% SLO satisfaction rate.

%% file: design.tex
\section{Design and Implementation}
\label{sec:design}


We propose \pname{}, a hardware-software co-design for component-level DVFS on NPUs.
\pname{} partitions the NPU into separate V/$f$ domains (\S\ref{sec:design:hw}).
It exposes fine-grained DVFS control by extending the NPU ISA (\S\ref{sec:design:isa}), which enables the compiler to control DVFS (\S\ref{sec:design:sw_dvfs}).
At runtime, the LLM serving system tracks request-level SLO slack and applies \pname{}'s DVFS decisions (\S\ref{sec:design:integration}).

\subsection{Hardware Support for Fine-Grained DVFS}
\label{sec:design:hw}

\begin{figure}[t]
    \centering
    \includegraphics[width=\linewidth]{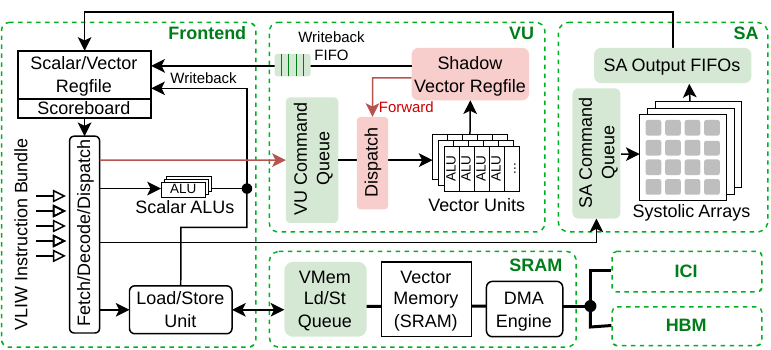}
    \caption{NPU core pipeline divided into different V/$f$ domains. Each V/$f$ domain is marked by a dashed \textcolor{mygreen}{green} box. The asynchronous FIFO queues used for inter-domain synchronization are also in \textcolor{mygreen}{green}. The queues for HBM and ICI controllers are omitted from the figure for simplicity.}
    \label{fig:npu_pipeline_vf_domains}
\end{figure}



\begin{figure*}[t]
    \centering
    \includegraphics[width=\linewidth]{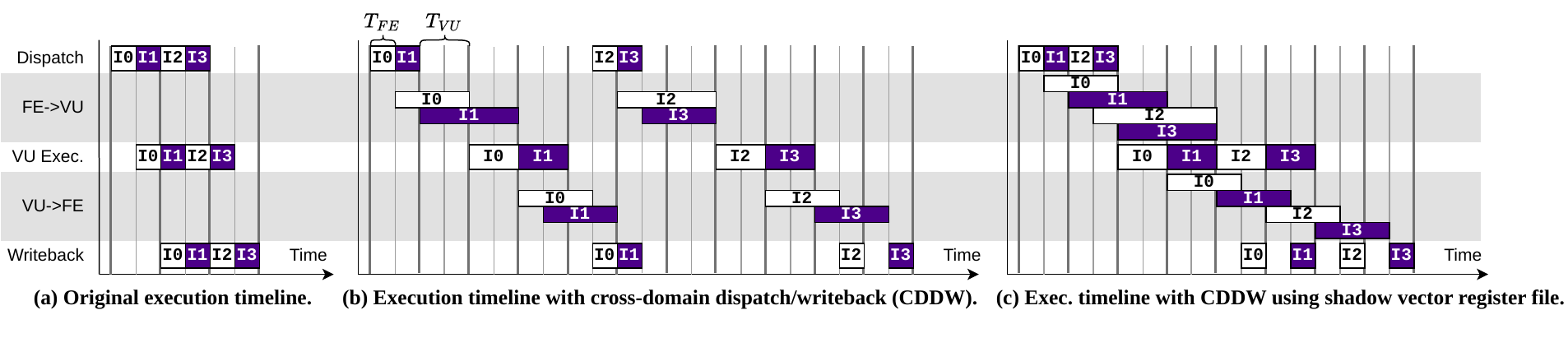}
    \caption{Example of instruction execution timeline with/without VU internal forwarding. VU frequency is set to the frontend frequency in (a), and halved in (b) and (c). Same-color commands have back-to-back data dependencies and must execute in order. \texttt{FE}: frontend. $\bm{T_{FE}}$: frontend clock period. $\bm{T_{VU}}$: VU clock period. For simplicity, domain crossing takes 1 cycle in each domain (i.e., only clock phase alignment is accounted for).}
    \label{fig:vu_forward_inst_timeline}
\end{figure*}

\begin{figure}[t]
    \centering
    \includegraphics[width=0.93\linewidth]{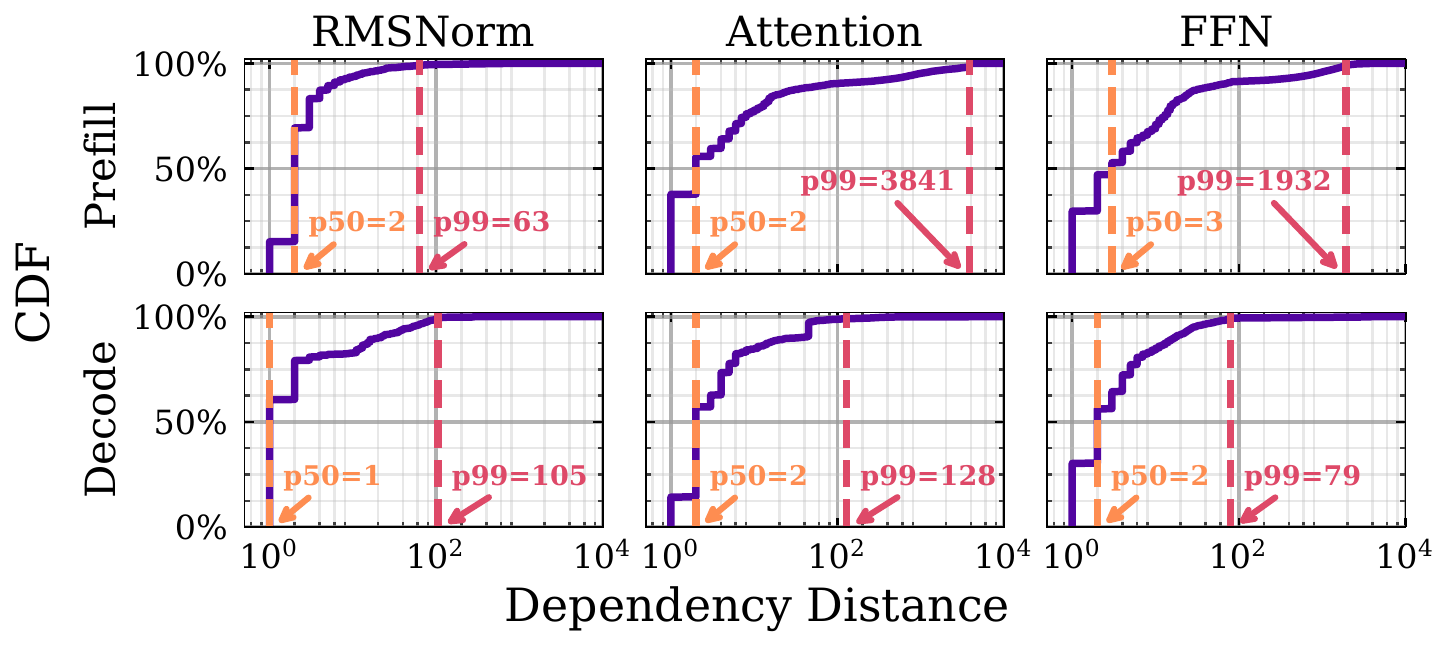}
    \caption{Dependency distances of VU commands in representative Llama3-70B layers, with input/output sequence length 4096/512, for the generated VLIW bundles on TPUv4 chips.}
    \label{fig:vu_inst_dependency_distance}
\end{figure}

In this section, we present the necessary hardware support for enabling component-level V/$f$ domains on NPUs.

\noindent
\textbf{Fine-grained V/$\bm{f}$ domains.}
\Cref{fig:npu_pipeline_vf_domains} shows the component-level V/$f$ domains in \pname{}.
Each domain has its own IVR and clock generator.
The \textit{frontend} domain coordinates the NPU execution. Its frequency is no lower than that of any other domain to avoid becoming a bottleneck.
The \textit{SA}, \textit{VU}, and \textit{SRAM} domains exchange data via the frontend's register file.

The \textit{HBM} and \textit{ICI} domains include on-chip controllers, I/O links, and DRAM dies.
While the controller is standard CMOS logic and supports DVFS by default, the V/$f$ of I/O and DRAM core are defined by strict standards~\cite{JESD235A2016}.
To abstract this complexity, \pname{} interfaces directly with the HBM/ICI controller, allowing software to set a target frequency state (see \S\ref{sec:design:isa}). The controller then utilizes an internal lookup table to select the appropriate V/$f$ for the I/O link or DRAM dies. 
This design ensures \pname{} seamlessly supports both current-generation hardware with controller-only DVFS and future standards that support different I/O voltages or DRAM core voltages, such as HBM4~\cite{hbm4}. 
\S\ref{sec:eval} evaluates both scenarios.


\noindent
\textbf{Cross-domain communications.}
For the SA, VU, and SRAM, \pname{} replaces the command queues, output FIFOs, and load/store queues with async FIFOs and voltage level shifters.
Each async FIFO must be sized appropriately to avoid pipeline bubbles at minimal hardware cost.
Let $T_i$ and $T_o$ be the clock periods of the input (producer) and output (consumer) domains.
To sustain maximum steady-state throughput, the FIFO must absorb the round-trip feedback latency of synchronizing the gray-coded read/write pointers via 3-stage flip-flops. Synchronizing the write pointer to the output domain takes $5T_o$ (1 cycle for phase alignment, 3 cycles for the flip-flops, and 1 cycle for logic). 
Similarly, synchronizing the read pointer to the input domain takes $5T_i$.
The round-trip latency is
\begin{equation}\label{eq:vf_crossing_delay}
    T_{rt} = 5T_i + 5T_o.
\end{equation}

Assuming the FIFO input/output rate is 1 entry/cycle (we can use multiple FIFOs to scale the aggregated bandwidth), the maximum steady-state pipeline throughput is $1/\max(T_i,T_o)$, and the minimum queue depth is
\begin{equation}
    Depth = \left\lceil T_{rt} \cdot \frac 1 {\max(T_i,T_o)} \right\rceil = \left\lceil \frac{5T_i + 5T_o}{\max(T_i,T_o)} \right\rceil.
\end{equation}
In the worst case, $Depth = 10$ when $T_i=T_o$.
Hence, each async FIFO in \pname{} has a minimum depth of 10 to prevent pipeline stalling due to cross-domain synchronizations.



\noindent
\textbf{VU internal forwarding.}
Many operators issue back-to-back data-dependent commands, especially elementwise and reduction operators on VUs; profiling shows that in at least 50\% of cases, data-dependent VU instructions are separated by at most 3 independent instructions.
\Cref{fig:vu_inst_dependency_distance} quantifies the dependency distances 
(i.e., distance is 1 for consecutive commands, and $n+1$ for those separated by $n$ commands) 
for all pairs of data-dependent VU commands in representative LLM layers. Only immediate (direct) dependencies are included.
With the V/$f$ domain crossing, 
the frontend scoreboard enforces that a consumer command cannot be dispatched until its producer retires, so
each command must wait for the complete round-trip latency of the previous command. 
This incurs pipeline bubbles that cannot be easily hidden via compiler optimizations.

As an example, \Cref{fig:vu_forward_inst_timeline}a shows the execution timeline of two pairs of VU commands (the dependency chain is \texttt{I0}$\to$\texttt{I2} and \texttt{I1}$\to$\texttt{I3}) with dependency distance 2 without V/$f$ domain crossing overhead.
Assuming VU frequency is halved, with domain crossing overhead ($\texttt{FE->VU}$ and $\texttt{VU->FE}$ in \Cref{fig:vu_forward_inst_timeline}b), the IPC (commands per frontend cycle) drops from 1 to 0.31. The compiler must find at least 6 independent VU commands to achieve the ideal IPC of 0.5.

To address this issue, we employ a lightweight \textit{shadow vector register file} in the VU domain so that register values can be forwarded to dependent VU commands 
without crossing between VU and frontend.
To avoid extra hardware scheduling logic, the $n$ shadow registers map to the first $n$ architectural registers.
\pname{} leverages the compiler to manage shadow registers.
The compiler calculates priority for register allocation based on dependency distance between each producer-consumer command pair. 
If the dependency distance between two commands is long enough to hide the FE-VU round-trip ($T_{rt}$ in \mbox{\Cref{eq:vf_crossing_delay}} plus VU execution delay), they can directly access the FE regfile without stalling, so shadow registers should not be allocated. 
Otherwise, the shorter the dependency distance is, the longer the FE would stall without forwarding,
so the compiler prioritizes command pairs with shortest dependency distances, assigning the first $n$ architectural registers when possible.

A local scoreboard in the VU domain enforces data dependencies that involve shadow registers. Hence, the frontend does not need to stall for VU commands that depend on them.
As shown in \Cref{fig:vu_forward_inst_timeline}c, this forwarding mechanism achieves ideal IPC without requiring extra independent commands, despite the longer pipeline warmup time.









\subsection{ISA Extension for \textmu s-scale DVFS on NPUs}
\label{sec:design:isa}

\begin{figure}[t]
    \centering
    \includegraphics[width=\linewidth]{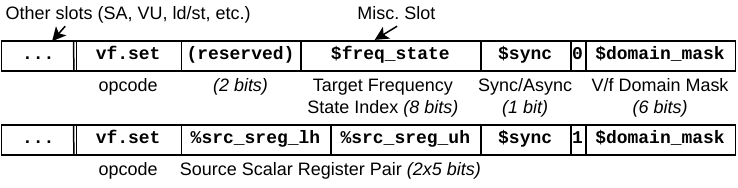}
    \caption{ISA extension for component-level DVFS.}
    \label{fig:dvfs_isa}
\end{figure}

As tensor operators execute on a $\mu$s-scale (see \Cref{tab:op_exe_time_cdf}), it is desirable to enable software DVFS configuration at sub-$\mu$s scale. 
To do so, we extend the NPU ISA with a new command ``\texttt{vf.set}'', as shown in \Cref{fig:dvfs_isa}.
The command occupies the ``misc.'' slot in a VLIW instruction and can be issued concurrently with SA, VU, and SRAM commands in other slots.
It offers two variants 
for
flexibility.

The first variant encodes target frequency as an 8-bit immediate value (\texttt{\$freq\_state}). This can encode up to 256 distinct frequency states, which is typically sufficient for fine-grained DVFS (in \S\ref{sec:impl}, we use 6 bits to encode 35 states from 0--1.7GHz at every 50MHz step).
\pname{} uses a lookup table for each component to map each frequency state to its minimum required voltage level.
\texttt{\$domain\_mask} specifies which components (frontend, SA, VU, SRAM, HBM, and ICI) to adjust the V/$f$ state. All selected components are scaled to the same target frequency, which is useful for setting all underutilized components in an operator to the lowest V/$f$ level or resetting them to the peak frequency.
The \texttt{\$sync} field specifies whether the \texttt{vf.set} is blocking. An asynchronous \texttt{vf.set} only blocks the target components, allowing other components to continue execution.

To allow independent V/$f$ adjustments of multiple components in one command, the second variant takes two 32-bit scalar registers as operands. The two registers are concatenated, and the lower 48 bits are taken as the target V/$f$ states of all domains.
Using register operands requires first loading the V/$f$ state values into scalar registers, which typically take an extra 1--3 scalar commands (like \texttt{lui} and \texttt{addi} in RISC-V).
As a VLIW bundle typically contains multiple scalar slots but only one misc. slot, this can save the number of bundles required compared to the first variant.
The bit fields for domains that are masked out are ignored by \texttt{vf.set}.


\subsection{Software-Managed DVFS}
\label{sec:design:sw_dvfs}


Given the SLO slack of an LLM request, \pname{} leverages the ML compiler to select an optimized DVFS plan. This requires navigating a combinatorial search space defined by hundreds to thousands of operators, 5 components to configure per operator, and 34 frequency choices -- a total of $(34^5)^{1000} \approx 10^{7657}$ DVFS plans.

To systematically navigate this search space, \pname{} relies on three key insights.
First, we can safely decouple the global search space into tractable \textit{operator-level} and \textit{request-level} search spaces.
Second, because the energy-delay trade-off is generally convex ($P \propto V^2f$), a greedy gradient descent works effectively to find optimized configurations. 
Third, as we can accurately calculate per-instruction domain crossing delay (see \S\ref{sec:design:hw}), we can build a cycle-accurate performance model based on the static instruction schedule within the compiler to rapidly assess 
a DVFS plan's execution time.

\noindent
\textbf{ML compiler workflow.}
The compiler takes the sequence of operators in an LLM request and its SLO slack as inputs and performs two passes of DVFS plan explorations.
Modern ML compilers like XLA~\mbox{\cite{xla}} and PyTorch~\mbox{\cite{pytorch2}} typically assume static computational graphs with known tensor shapes and bounded loops. This enables accurate cost models and allows the compiler to perform aggressive compilation-time optimizations, such as tiling, operator fusion, and memory layout selection, based on the tensor shapes.


\mbox{\pname{}} performs the DVFS plan search after the compiler has finished all optimizations and generated the final VLIW instruction schedule for each operator.
For each operator, \mbox{\pname{}} generates DVFS plans that closely estimate the Pareto frontier of energy-delay tradeoffs.
Then, it iteratively picks an optimized DVFS plan from the Pareto frontier of each operator until exhausting the SLO slack and generates the NPU program with \texttt{vf.set} commands.

\newcommand\mycommfont[1]{\footnotesize\sffamily\textcolor{purple}{#1}}
\SetCommentSty{mycommfont}
\begin{algorithm}[t]
\small
\DontPrintSemicolon
\SetAlgoNoEnd
\SetAlgoLined
\SetKwFunction{FMain}{{\it get\_pareto\_plans\_for\_op}}
\SetKwProg{Fn}{Function}{:}{}

\Fn{\FMain{op}}
{
  \tcp{\color{purple}{start with lowest frequencies (50MHz) for all components}}
  \textit{cur\_plan} $\gets$ $(50, 50, 50, 50, 50)$ \;
  \textit{explored\_plans} $\gets$ \{ \textit{cur\_plan} \} \;
  
  \While{any(c $<$ MAX\_FREQ for c in cur\_plan)}
  {
    \tcp{\color{purple}{increment freq of $\ge 1$ components (up to 31 plans)}}
    \textit{candidates} $\gets$ \textit{generate\_next\_candidates(cur\_plan)} \;
    \textit{best\_cand} $\gets$ \textbf{null} \;
    \textit{best\_gradient} $\gets$ $\infty$ \;
    
    \For{cand in candidates}
    {
      \tcp{\color{purple}{evaluate candidates using cost model}}
      $\Delta E$, $\Delta T$ $\gets$ \textit{evaluate\_diff(cur\_plan, cand)} \;
      
      \tcp{\color{purple}{pick plan with the best energy-delay tradeoff}}
      \If{$\Delta T < 0$ and $(\Delta E / |\Delta T|) < best\_gradient$}
      {
        \textit{best\_gradient} $\gets$ $\Delta E / |\Delta T|$ \;
        \textit{best\_cand} $\gets$ \textit{cand} \;
      }
    }
    
    \tcp{\color{purple}{stop if no candidate achieves better performance}}
    \If{best\_cand is \textbf{null}}{
      \textbf{break} \;
    }
    
    \textit{cur\_plan} $\gets$ \textit{best\_cand} \;
    \textit{explored\_plans} $\gets$ \textit{explored\_plans} $\cup$ \{ \textit{cur\_plan} \} \;
  }
  
  \tcp{\color{purple}{post-processing pass to filter out non-Pareto plans}}
  \textit{pareto\_plans} $\gets$ \textit{filter\_pareto\_plans(explored\_plans)} \;
  \Return{pareto\_plans}
}

\caption{Operator-level DVFS plan exploration.}
\label{alg:dvfs_intra_op}
\end{algorithm}

\noindent
\textbf{Operator-level exploration.}
To perform static instruction scheduling, NPU compilers typically use 
an instruction-level cost model, which takes
instruction latencies and a sequence of VLIW bundles
to evaluate the execution time of an operator.
DMA/RDMA operations for HBM/ICI use a well-established link model~\cite{ahead:2019,loggp:1995}.
\pname{} reuses this cost model 
with
new instruction latencies determined by the frequency setting and domain-crossing delay (Equation \ref{eq:vf_crossing_delay}).

For each operator, there are $34^5\approx 45$M possible DVFS plans.
\pname{} uses gradient descent to identify the (approximate) Pareto plans that represent the optimal energy-delay tradeoffs for this operator, as shown in \Cref{alg:dvfs_intra_op}.
It starts with the lowest frequencies for all components, which represent the plan with the largest execution time (line 2).
In each iteration, it generates the next set of candidate plans by incrementing the frequency of one or more components to the next available frequency point, yielding up to $2^5 - 1=31$ plans (line 5).
The candidate plans are evaluated with the cost model, and \pname{} picks the next plan to be the one that achieves the best tradeoff between energy and execution time (i.e., the plan with the smallest $\Delta E / |\Delta T|$ where $\Delta E$ and $\Delta T$ are the differences in energy consumption and execution time between the current and candidate plans) (line 8--12).
The algorithm stops when no candidate plans achieve better performance or all components reach peak frequency.
The selected plan in each iteration is recorded, and a final post-processing pass filters out Pareto plans from the explored plans (line 17).
In the worst case, the greedy search increments only one component's frequency per iteration, so the maximum number of evaluated plans per operator is $5 \times 34 \times (2^5-1)=5270$.



\noindent
\textbf{Request-level exploration.}
Given the Pareto plans for all operators, \pname{} iteratively searches for the request-level DVFS plan that minimizes energy consumption without violating SLO, using a similar heuristic based on the $\Delta E / \Delta T$ gradient.
\pname{} starts with the fastest Pareto plan for each operator.
In each iteration, it picks the operator whose next plan on the Pareto frontier saves the most energy while incurring the least performance degradation and updates the plan.
The algorithm stops if either the SLO slack or candidate plans are exhausted.
In the worst case, the request-level search takes $O(NP)$ iterations, where $N$ is the number of operators and $P$ is the number of Pareto plans per operator.






\subsection{LLM Serving System Integration}
\label{sec:design:integration}

\noindent
\textbf{Offline DVFS plan generation.}
In production LLM serving systems, multiple graphs are compiled ahead-of-time before deployment for various pre-bucketed batch sizes and sequence lengths.
At runtime, the request scheduler~\cite{vllm:sosp23} selects the most appropriate pre-compiled graph to execute.
The batch size and sequence length are padded to the smallest bucket that can accommodate them, avoiding runtime compilation overhead. Background compilation of extra buckets can be triggered if an unseen batch size or sequence length becomes frequent.
\pname{} additionally generates the DVFS plan for each batch size, sequence length, and SLO slack.
The generated results can be shared by all services using the same backbone LLM.
By default, \mbox{\pname{}} compiles graphs for all power-of-2 batch sizes from 1 to 16 and multiple-of-16 batch sizes beyond 16, up to 128. For sequence lengths, it uses 128-token increments up to 1024 tokens and doubles the increment at each subsequent power-of-2 boundary, e.g., 256-token increments up to 2048 and 512-token increments up to 4096~\mbox{\cite{vllm:sosp23}}. For each graph, \mbox{\pname{}} covers an SLO slack of 0\%, 2\%, 5\%, and 10\%, which are sufficient for achieving most DVFS benefits (see \mbox{\Cref{fig:eval_request_all}}). This yields at most 1728/2112/2304 compilations for up to 128K/512K/1M sequence length. The compilation time is manageable, as this can be done offline just once with massive CPU cores (see \mbox{\S\ref{sec:eval:search_space}}).


\noindent
\textbf{DVFS-aware request scheduling.}
At runtime, the request scheduler picks requests from a queue to form a batch~\cite{vllm:sosp23,sglang:nips24}.
\pname{} preserves the batching policy of the request scheduler, which is typically designed to maintain SLO and improve system throughput.
\pname{} tracks each request's queuing delay, $T_q$, and computes its target execution time $T_{target}=SLO-T_q$.
After forming a batch, \pname{} looks up the best DVFS plan based on the batch size, sequence length, and the smallest $T_{target}$ in the batch.
To ensure SLO for all requests in a batch, \pname{} selects the DVFS plan assuming all requests have the same sequence length as the longest request.


Slowing down the execution of a batch to save energy may exacerbate head-of-line blocking for subsequent requests.
When the system is stressed by load spikes or during peak hours, slowing down the execution causes the entire system throughput to drop proportionally, which can cause significant SLO violations.
To avoid this, \pname{} enforces two safeguard rules. 
First, it adjusts the $T_{target}$ of a batch such that it will not directly cause any request in the queue to violate SLO.
Second, it monitors the SLO satisfaction rate in the past time window (5 minutes by default) and temporarily locks the NPU chips at the peak frequency if the SLO drops under a specific target (e.g., more than 1\% of requests violate SLO).

\noindent
\textbf{Support for custom operators.}
By default, modern ML compilers like XLA~\cite{xla} and PyTorch~\cite{pytorch2} assume static computational graphs.
For MoE models, dynamic behavior (e.g., varying token counts for MoE experts) is often implemented as a custom operator with a custom cost model.
The ML compiler can treat this custom operator similarly to other compiler-generated ones during cost-model-based optimizations.
For example, the execution time of an MoE expert depends on the number of tokens it processes.
By default, the cost model uses an expert capacity factor~\mbox{\cite{moe_imbalance:iclr26}} to estimate the worst-case token count for an expert\footnote{The expert capacity factor measures the ratio between the token count for the most loaded expert and the average token count for all experts.}, and \mbox{\pname{}} conservatively selects DVFS plans that do not violate SLO, despite sacrificing potential DVFS opportunities. The user can adjust the expert capacity factor based on real workload profiling and SLO requirements.






\subsection{Implementation}
\label{sec:impl}

\noindent
\textbf{Hardware synthesis.}
To verify 
\pname{}'s pipeline design (\S\ref{sec:design:hw}), we implement our design based on the open-source Coral NPU core~\cite{coralnpu}, which is an in-order 4-issue RISC-V core with a RISC-V-compliant vector core (VU), a custom $16\times 16$ matrix core (SA), and SRAM. We implement \texttt{vf.set} (\S\ref{sec:design:isa}) as a custom RISC-V control instruction and add the async FIFOs and shadow vector register file to the pipeline. We use the unified power format (UPF), a standard format to specify power-related settings in EDA tools~\mbox{\cite{upf_ieee,chipverify_upf_intro}}, to define voltage islands for the frontend, SA, VU, and SRAM. We synthesized and placed \& routed the pipeline in Synopsys with ASAP7 PDK~\cite{asap7}.
For async FIFOs, we use the default queue size $+10$ for each component based on our analysis in \S\ref{sec:design:hw}.
\Cref{tab:hardware_area_power} lists the area/power breakdown. \mbox{\pname{}} incurs 6.16\% area overhead in total: \mbox{\pname{}}'s microarchitecture and voltage level shifters incur 4.58\%/5.45\% area/power overhead for the Coral NPU core; the DVFS-enabled IVR incurs 1.58\% area overhead compared to a fixed-voltage IVR\footnote{Even without DVFS, a fixed-voltage IVR is necessary for mitigating transient voltage droops and improving off-chip power delivery efficiency.}.

The overhead will be relatively smaller for larger NPU chips whose SAs/VUs and their FIFO sizes are larger.
The IVR area scales with the chip's peak power, which is typically manageable for modern datacenter AI chips~\mbox{\cite{dvfs_gpu:amd:asplos23}}.
As projected in \mbox{\Cref{tab:hardware_area_scale}}, for a TPUv4/TPUv5p chip~\mbox{\cite{tpuv4:isca23,tpuv5p}}, the total area overhead is 2.20\%/2.31\%, with 0.46\%/0.4\% due to \mbox{\pname{}}'s structures, 0.4\%/0.37\% due to voltage level shifters, and 1.34\%/1.54\% due to the DVFS-enabled IVR.

\begin{table}[t]
  \centering
  \caption{Area and power overhead breakdown of \pname{} for the Coral NPU core prototype~\mbox{\cite{coralnpu}}. All percentages are normalized to the total area of the original Coral NPU core.}
  \label{tab:hardware_area_power}
  \footnotesize
  \resizebox{\columnwidth}{!}{%
  \begin{tabular}{l|rr|r}
      \toprule
      & \multicolumn{2}{c|}{\textbf{Area}} & \multicolumn{1}{c}{\textbf{Power (mW)}} \\
      & \textbf{($\mu m^2$)} & \textbf{(\%)} & \multicolumn{1}{c}{@ 0.7V, 500MHz} \\
      \midrule
      Frontend & 11,853.0 & 4.07\% & 3.287 \\
      Systolic Array ($16\times 16$) & 214,850.4 & 73.74\% & 127.074 \\
      Vector Unit (128-bit) & 46,251.2 & 15.87\% & 14.509 \\
      SRAM (32KB) & 11,007.5 & 3.78\% & 1.346 \\
      \midrule
      FE $\leftrightarrow$ SA Sync FIFO (16 Entries) &  2,790.7 & 0.96\% & 2.002 \\
      FE $\leftrightarrow$ VU Sync FIFO (16 Entries) & 3,337.4 & 1.15\% & 2.394 \\
      FE $\leftrightarrow$ SRAM Sync FIFO (4 Entries) & 1,287.2 & 0.44\% & 3.683 \\
      \midrule
      \textbf{Original NPU Core Total} & \textbf{291,377.4} & \textbf{100.0\%} & \textbf{154.295} \\
      \textbf{Fixed-Voltage IVR~\cite{Zelikson2023DigitalLDO}} & \textbf{1,800} & \textbf{0.62\%} & - \\
      \midrule
      FE $\leftrightarrow$ SA Async FIFO  (26 Entries) & 4,570.6 & 1.57\% & 2.007 \\
      FE $\leftrightarrow$ VU Async FIFO  (26 Entries) & 5,465.6 & 1.88\% & 2.400 \\
      FE $\leftrightarrow$ SRAM Async FIFO  (14 Entries) & 4,540.7 & 1.56\% & 3.692 \\
      Shadow Vector Regfile (4 Regs) & 261.1 & 0.09\% & 0.747 \\
      Voltage Level Shifters (FE-SA) & 1858.5 & 0.64\% & 2.40 \\
      Voltage Level Shifters (FE-VU) & 2222.4 & 0.76\% & 2.86 \\
      Voltage Level Shifters (FE-SRAM) & 1846.3 & 0.63\% & 2.38 \\
      \midrule
      \textbf{eNPU Core Total} & \textbf{304,727.3} & \textbf{104.58\%} & \textbf{162.702} \\
      \textbf{DVFS-Enabled IVR~\cite{dvfs_fivr:2016,dvfs_gpu:amd:asplos23}} & \textbf{6418.5} & \textbf{2.20\%} & - \\
      \bottomrule
  \end{tabular}%
  }
\end{table}

{
\begin{table}[t]
    \centering
    \caption{Projected area overhead of {\pname{}} on real TPU chips. The SA/VU sync FIFO is per SA/VU. The SRAM sync FIFO is per chip and scales with the number of read/write (R/W) ports. Each FIFO entry has the same size as a vector register. *TPUv5p die area is not publicly disclosed, so we use NVIDIA A100 GPU's die area~\mbox{\cite{a100}} as a proxy, as this is one of the biggest 7nm dies given the reticle size limit.}
    \label{tab:hardware_area_scale}
    \footnotesize
    \resizebox{\columnwidth}{!}{%
    \begin{tabular}{l rrrr}
        \toprule
        \textbf{Parameter} & \multicolumn{2}{c}{\textbf{TPUv4}} & \multicolumn{2}{c}{\textbf{TPUv5p}} \\
        \midrule
        \# of SAs/VUs/SRAM Ports & \multicolumn{2}{c}{8/4/4} & \multicolumn{2}{c}{8/6/6} \\
        Vector Register Size & \multicolumn{2}{c}{4 KB} & \multicolumn{2}{c}{4 KB} \\
        SA/VU/SRAM Sync FIFO Depth & \multicolumn{2}{c}{16/16/16} & \multicolumn{2}{c}{16/16/24} \\
        SA/VU/SRAM Async FIFO Depth & \multicolumn{2}{c}{26/26/26} & \multicolumn{2}{c}{26/26/34} \\
        \midrule
         & \textbf{(mm$^2$)} & \textbf{\%} & \textbf{(mm$^2$)} & \textbf{\%} \\
        \midrule
        \textbf{Original Total Die Area} & \textbf{600} & \textbf{100\%} & \textbf{*826} & \textbf{100\%} \\
        \midrule
        FE $\leftrightarrow$ SA Sync FIFO & 2.405640   & 0.40\% & 2.405640   & 0.29\% \\
        FE $\leftrightarrow$ VU Sync FIFO & 1.202820   & 0.20\% &1.804230 & 0.22\% \\
        FE $\leftrightarrow$ SRAM Sync FIFO & 0.601410  & 0.10\% & 1.353172 & 0.16\% \\
        Fixed-Voltage IVR~\cite{Zelikson2023DigitalLDO} & 3.27  & 0.54\% & 5.16  & 0.62\% \\
        \midrule
        FE $\leftrightarrow$ SA Async FIFO &3.939825   &  0.66\%& 3.939825   & 0.48\% \\
        FE $\leftrightarrow$ VU Async FIFO & 1.969912  &   0.33\%& 2.954869  & 0.36\% \\
        FE $\leftrightarrow$ SRAM Async FIFO & 0.984956  & 0.16\% & 1.932029  & 0.23\% \\
        Shadow Vector Regfile (4 Regs/Core) & 0.075176   & 0.01\% &0.075176  & 0.01\% \\
         Voltage Level Shifters (FE-SA) & 1.379705     & 0.23\% & 1.379705    & 0.17\% \\
         Voltage Level Shifters (FE-VU) & 0.689853     & 0.11\% & 1.034779  & 0.13\% \\
         Voltage Level Shifters (FE-SRAM) & 0.344926     & 0.06\% & 0.676586  & 0.08\% \\
        DVFS-Enabled IVR~\cite{dvfs_fivr:2016,dvfs_gpu:amd:asplos23} & 11.3 & 1.88\% & 17.82 & 2.16\% \\
        \midrule
        \textbf{Total Area Overhead} & \textbf{13.204} & \textbf{2.20\%} & \textbf{19.090} & \textbf{2.31\%} \\
        \bottomrule
    \end{tabular}%
    }
\end{table}
}
\begin{figure}[t]
    \centering
    \includegraphics[width=\linewidth]{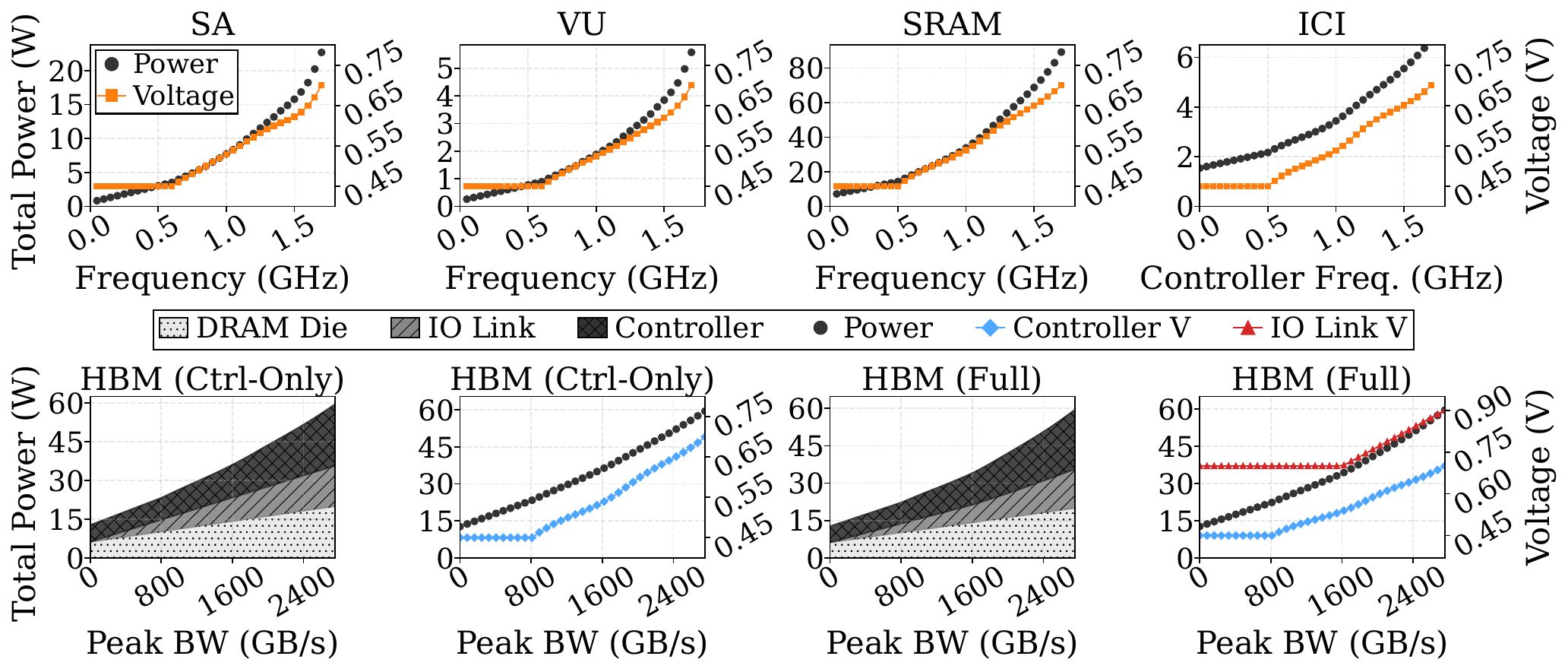}
    \caption{Power usage and minimum required voltage of components at various performance levels (peak frequency or bandwidth). The SA/VU plot shows the power of a single SA/VU. The SRAM plot uses 128MB SRAM with six read/write ports. For HBM, the total power includes on-chip controller, I/O, and HBM die. For ICI, the total power includes controller and I/O. For all plots, we show the power at peak component utilization (i.e., peak achievable FLOPs/sec or bandwidth).}
    \label{fig:synth_peak_power}
\end{figure}

\noindent
\textbf{Simulator methodology.}
We build our simulator framework to model \pname{} at instruction, request, and LLM service levels. 

First, we build an instruction-level performance model to estimate the impact of \pname{}'s NPU pipeline changes.
It takes the VLIW assemblies for each operator (dumped from the XLA compiler in a Cloud TPUv4 VM~\cite{dump_tpu_vliw}), reconstructs the instruction dependencies, and estimates the execution time under different V/$f$ settings by accounting for frequency-dependent instruction latencies, async FIFO delays, and VU internal forwarding (see \S\ref{sec:design:hw}).
As the TPU compiler is proprietary, we build our own instruction scheduling passes (including loop unrolling and software pipelining) based on the reconstructed instruction dependency graph and implement \pname{}'s DVFS algorithm (\S\ref{sec:design:sw_dvfs}). 
We validate this performance model on different operator types on a TPUv4 chip.

For request-level simulation, we use a production-level NPU chip simulator that simulates each component (SA, VU, SRAM, ICI, and HBM) at tile level. 
Taking the model graph as input, it computes each operator's per-component execution time using the tile shape, FLOPs per tile, and component-specific parameters (e.g., SA/VU width, HBM/ICI bandwidth), plus SRAM, HBM, and ICI traffic.
The per-component execution time is validated against real TPU chips across operator types and tensor shapes.
The simulator scales execution time and power of each component based on the selected V/$f$,
models V/$f$ transitioning delay and IVR power conversion loss, and invokes the instruction-level model to quantify the impact of domain-crossing delay and shadow regfile size.

We combine the performance simulator with a power model to derive energy consumption.
As TPUs lack a publicly available power-measurement API, 
we build our own power model based on RTL prototypes of SA, VU, and SRAM, plus public datasheets for HBM~\cite{hbm_ip:amd,hbm_ip:atria,hbm_ip:rambus,hbm4} and Google's datacenter optical network~\cite{Jupiter}.
We use Synopsys to synthesize our components with ASAP7~\cite{asap7} using the nominal 0.7V supply voltage and a 1.7GHz frequency target (to roughly match TPUv5p).
We use Synopsys PrimeTime to estimate static and dynamic power at different V/$f$ corners (see \Cref{fig:synth_peak_power}).
For HBM, we include two power models: ``HBM (Ctrl-Only)'' fixes the voltage of I/O link and DRAM dies and only scales the V/$f$ of on-chip memory controller. This resembles current HBM technology that requires fixed I/O and DRAM core voltages for stability~\cite{JESD235A2016}. ``HBM (Full)'' adjusts V/$f$ for both controller and I/O link, reflecting future HBM4 standards~\cite{hbm4} that allow multiple I/O voltage levels.
For ICI, we only apply DVFS to the controller, as DVFS cannot apply to the optical link.
We validated our power model against published data~\cite{neurometer:hpca21,tpuv4:isca23,regate:micro25}: the estimated TDP of a TPUv3/TPUv4 chip is within 4\%/8\%.

To evaluate \pname{} at LLM service level, we use the request-level simulator as backend to build a cluster simulator, which replays LLM inference service traces and evaluates latency, end-to-end throughput, and SLO satisfaction rate. It implements the same request routing and continuous batching policies as a production LLM serving engine~\cite{vllm:sosp23}. It also models the KV cache transfer between prefill/decode instances~\cite{DistServe:osdi24,splitwise:isca24}.
We validated the prefill/decode throughput and the TTFT/TPOT of the cluster simulator against vLLM deployment~\cite{vllm:sosp23} on a Cloud TPUv4-64 instance.

%% file: evaluation.tex
\section{Evaluation}
\label{sec:eval}

{
\begin{table}[t]
    \centering
    \caption{NPU chip and IVR specifications used in our simulations. The NPU chip data are based on TPUv5p~\cite{regate:micro25}. The IVR data assume a hybrid 22 nm SCVR/DLDO design~\cite{dvfs_fivr:2016,dvfs_gpu:amd:asplos23}.}
    \label{tab:npu_specs}
    \footnotesize
    \resizebox{\columnwidth}{!}{%
    \begin{tabular}{ll}
        \toprule
        \textbf{Parameter} & \textbf{Value} \\
        \midrule
        TFLOPS (bf16) & 445.6 TFLOP/s @ 1.7 GHz \\
        \# of SAs/VUs & 8 SAs \& 6 VUs, 32 architectural vector registers \\
        SA Config. & 128$\times$128 (bf16 MAC) \\
        VU Config. & 8$\times$128 SIMD ALU, 4 shadow vector registers \\
        SRAM & 128 MB, six 128$\times$8$\times$4B read/write ports \\
        HBM & 95 GB, 2765 GB/s \\
        ICI & 3D torus, 6 links/chip, 100 GB/s/link \\
        DCN & 50 GB/s/chip \\
        \midrule
        IVR Transition Delay & 4 ns / 54 mV \\
        $V_{in}$ $\to$ Min/Max $V_{out}$ & 0.75 V $\to$ 0.45 V / 0.7 V \\
        Power Conversion Eff. & SCVR: 63\%@0.45V; DLDO: 44\%@0.45V--87\%@0.7V \\
        Clock Frequency & 0 (clock-gated)--1.7 GHz, 50 MHz steps \\
        \bottomrule
    \end{tabular}%
    }
\end{table}
}

\begin{figure*}[t]
    \centering
    \includegraphics[width=\linewidth]{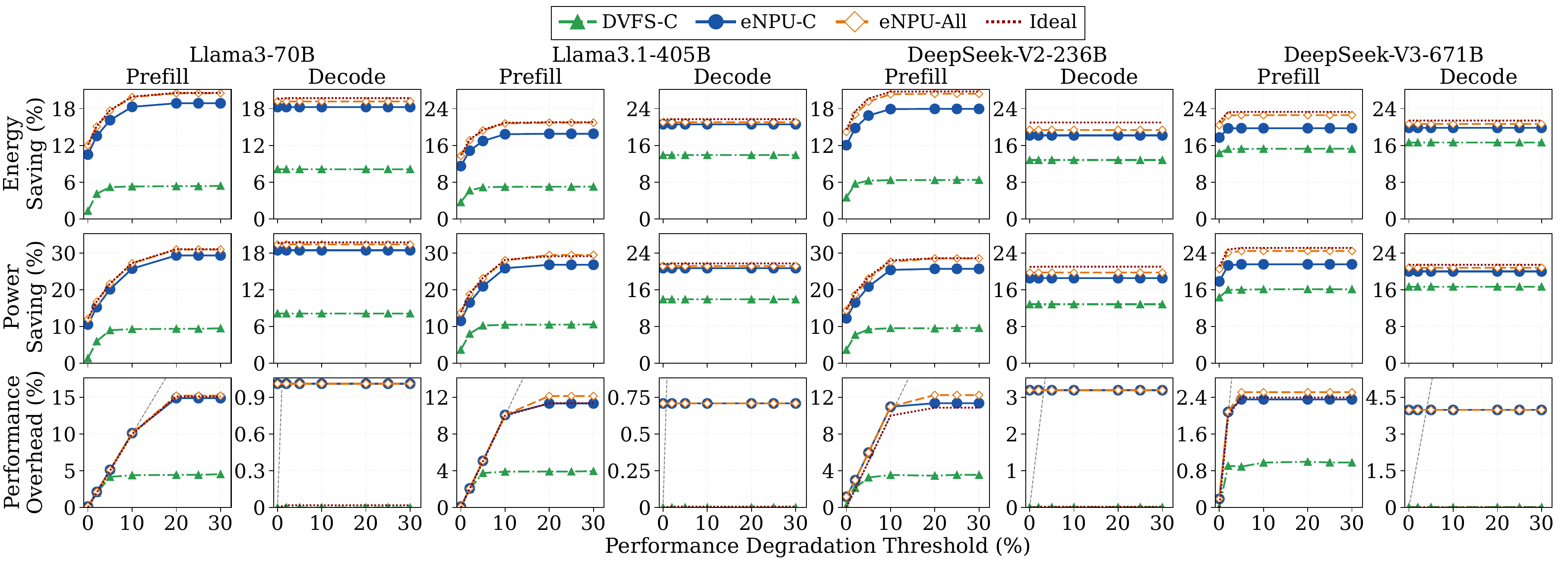}
    \caption{Energy saving (top), power saving (middle), and performance overhead (bottom) of an LLM request, given different SLO slacks (i.e., performance degradation thresholds).}
    \label{fig:eval_request_all}
\end{figure*}

We show that:
(1) At 0\% performance degradation threshold, \pname{} saves 15.1\%/20.1\% energy for LLM prefill/decode. As the threshold increases to 10\%, its energy savings rise to 20.9\%/20.1\%. \pname{} achieves 9.4\%--12.6\% (prefill) and 7.2\% (decode) higher energy savings than the state-of-the-art NPU DVFS solution on average (\S\ref{sec:eval:request_energy}).
(2) \pname{}'s compiler-driven DVFS algorithm efficiently navigates the huge search space (\S\ref{sec:eval:search_space}).
We also analyze \pname{}'s overhead (\S\ref{sec:eval:overhead}) and its sensitivity to the shadow vector register file size (\S\ref{sec:eval:vu_ifd}).
(3) \mbox{\pname{}}'s benefits generalize to different temporal DVFS granularities, V/$f$ domain count, sequence lengths, and expert load imbalances in MoE models (\S\ref{sec:eval:sens}).
(4) Combining DVFS and power gating can further achieve up to 24.9\%/32.4\% energy savings for prefill/decode (\S\ref{sec:eval:dvfs_pg}).
(5) For end-to-end LLM services, \pname{} achieves 16.3\%--20.5\% energy savings with strict SLO guarantees (\S\ref{sec:eval:llm_service}). 

\subsection{Experimental Setup}
\label{sec:eval:setup}

\begin{figure}[t]
    \centering
    \includegraphics[width=\linewidth]{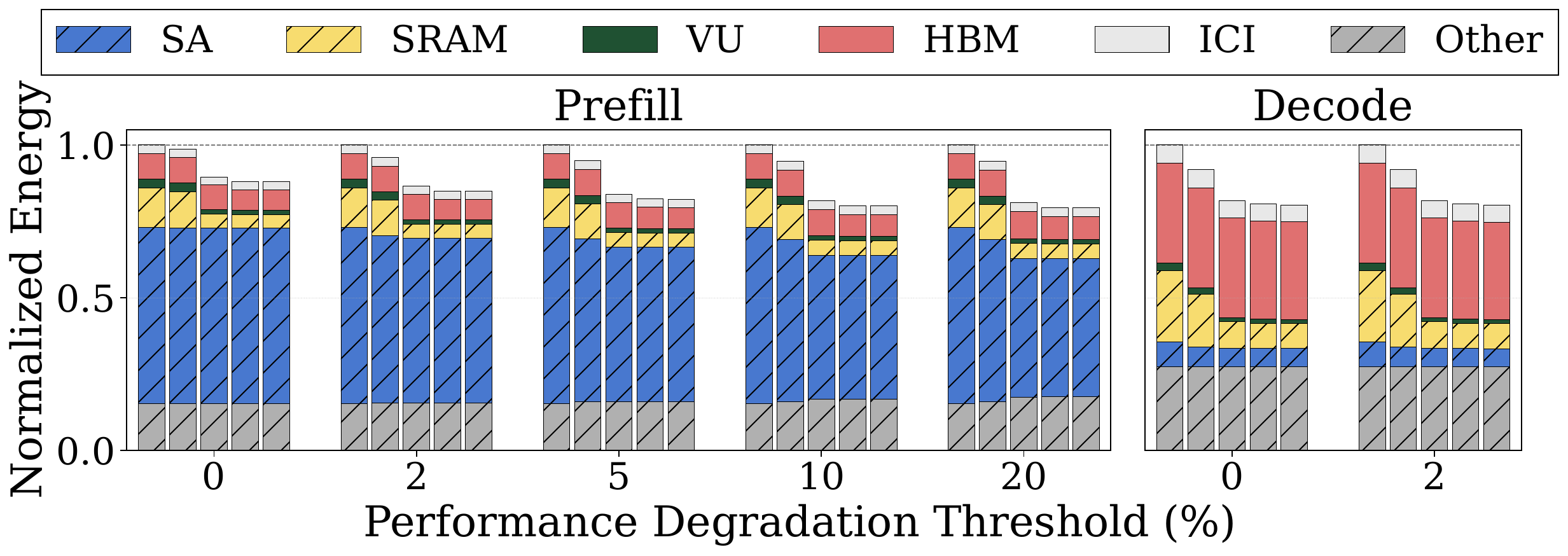}
    \caption{Component-level breakdown of energy savings in an LLM request. The bars from left to right are: \nodvfs, \dvfsc, \pname{}-C, \pname{}-All, and \ideal. Due to space limitations, we show the breakdown of Llama3-70B as an example. We only show 0\% and 2\% performance degradation thresholds for decode, as the energy savings already saturate at 2\%.}
    \label{fig:eval_energy_breakdown_component}
\end{figure}

\noindent
\textbf{Workloads.}
We evaluate four LLMs, including dense (Llama) and MoE (DeepSeek) models. We study prefill and decode separately since they exhibit distinct resource demands, following state-of-the-art prefill/decode disaggregation (PDD)~\cite{DistServe:osdi24,splitwise:isca24}.
We evaluate both single-request (\S\ref{sec:eval:request_energy}--\S\ref{sec:eval:dvfs_pg}) and LLM service-level (\S\ref{sec:eval:llm_service}) benefits.
For single-request evaluation, we use 4096/512 input/output sequence lengths and study different performance degradation thresholds (i.e., SLO slacks). We study other sequence lengths in \mbox{\S\ref{sec:eval:sens}}. 
For service-level evaluation, we replay production traces~\cite{dynamollm:hpca25} (\Cref{tab:slo_setting} and \Cref{tab:pd_node_sweep} for the SLO targets and cluster setup).
As the trace only contains timestamps and input/output sequence lengths per request with no prompt or output content, we simulate the worst-case expert capacity factor~\mbox{\cite{moe_imbalance:iclr26}} for MoE models by default. We study different capacity factors in \mbox{\S\ref{sec:eval:sens}}.

\noindent
\textbf{Simulator setup.}
We use TPUv5p~\cite{tpuv5p} specifications (see \Cref{tab:npu_specs}).
We use a modern IVR~\cite{dvfs_fivr:2016,dvfs_gpu:amd:asplos23} that achieves high power conversion efficiency and low V/$f$ scaling delay.

\noindent
\textbf{Baselines.}
We compare five designs:
\begin{itemize}[leftmargin=*]
    \item \textbf{\nodvfs}: the NPU chip always runs at peak V/$f$.
    It employs a fixed-voltage IVR with an 89\%--93\% power conversion efficiency~\mbox{\cite{Zelikson2023DigitalLDO}}.
    It does not incur domain crossing overhead.

    \item \textbf{\dvfsc}~\cite{ascend_dvfs:asplos25}: the state-of-the-art NPU DVFS solution, which scales the V/$f$ of the compute domain (SA, VU, and SRAM). 
    To ensure fair comparison, we use the ns-scale IVR transition delay in \mbox{\Cref{tab:npu_specs}} and modify the design's genetic algorithm to search for the optimal compute domain frequency per operator, instead of per ms-scale epoch as the original paper proposed.
    We study the impact of temporal DVFS granularity in {\S\ref{sec:eval:sens}}.

    \item  \textbf{\base}: our design with DVFS for HBM controller but not I/O and DRAM dies.

    \item  \textbf{\full}: our design with full DVFS for both HBM controller and I/O link.

    \item  \textbf{\ideal}: this design uses exhaustive search for operator-level and greedy search for request-level exploration 
    (as sweeping all request-level plans is impossible).
    It still accounts for IVR power conversion loss, but no V/$f$ scaling delay or domain crossing overhead. 

\end{itemize}

\subsection{Energy and Power Savings}
\label{sec:eval:request_energy}


We first analyze the energy/power savings for a single LLM request with input/output sequence length 4096/512, as shown in \Cref{fig:eval_request_all}.
We study various performance degradation thresholds (i.e., SLO slacks).
As the power-saving trend is similar to that of energy savings, we focus on the energy savings in the following discussion.

\noindent
\textbf{Overall energy savings.}
With the entire compute core (SAs, VUs, and SRAM) as a single V/$f$ domain, \dvfsc achieves 5.4\%--15.3\% energy savings for prefill and 8.1\%--16.7\% for decode.
\dvfsc's energy saving at low SLO slacks is limited, 
as it must set a uniform frequency for all components, forcing them to follow the bottleneck component's frequency.
In contrast, \base precisely throttles each component to its optimal frequency.
On average, it achieves 7.2\%/6.4\% (prefill/decode) higher energy savings than \dvfsc at 0 SLO slack and 10.3\% (prefill) at large SLO slacks.
\full achieves slightly better energy savings than \base by allowing DVFS for HBM I/O voltage, and it closely tracks 
\ideal.
The limited extra saving of \full is due to the fixed DRAM core voltage in current HBM generations; supporting DRAM-core DVFS 
could further improve memory energy efficiency.


\noindent
\textbf{Prefill vs. decode.}
For prefill, most designs reach maximum energy savings at $\sim$10\% SLO slack since prefill is compute-bound and offers headroom to trade performance for energy by throttling SA, VU, and SRAM.
As their power scales super-linearly with frequency ($P \propto V^2f$), additional slack enables further energy savings.
In contrast, decode reaches maximum energy savings at 0--2\% SLO slack since decode is memory-bound and HBM dominates dynamic energy (see \Cref{fig:eval_energy_breakdown_component}).
The HBM subsystem has a near-linear power-frequency relationship (see \Cref{fig:synth_peak_power}), so lowering frequency reduces power only near-linearly and extends execution time linearly, 
yielding minimal net energy reduction while increasing static energy for other components.
Thus, most decode energy savings come from throttling non-bottlenecking compute components (SA, VU, and SRAM) to reduce static power.

\subsection{Analysis of \pname{}'s DVFS Plan Search}
\label{sec:eval:search_space}




\noindent\textbf{Operator-level search.}
\Cref{tab:num_explored_dvfs_plans} shows how \pname{} tackles the combinatorial search space of component-level DVFS. Because \dvfsc restricts its search space to a coarse-grained V/$f$ domain (34 V/$f$ choices per operator), it only evaluates 578--1,394 total operator-level plans across unique operators 
(LLMs repeat many operators across stacked transformer layers).
\pname{}'s component-level DVFS unlocks $34^5 \approx 45M$ possibilities per operator; its two-level gradient descent 
explores 15K--38K plans to discover near-Pareto ones.

\noindent\textbf{Request-level search.}
\Cref{tab:num_explored_dvfs_plans} also compares request-level search overhead.
We empirically configure \dvfsc's genetic algorithm to use 500 generations and population size 1000 per performance degradation threshold, 
seeding larger thresholds with results from smaller thresholds to speed up convergence
(e.g., results of 2\% are used as the seed for 5\%).
This forces it to explore 2.5M request-level plans total, regardless of the model's operator count.
In contrast, \pname{} converges to a near-optimal plan within only 2.9K--119K evaluations. 
Notably, decode has a limited request-level search space.
As degrading decode performance often yields no extra energy savings (see \S\ref{sec:eval:request_energy}), there is usually only one Pareto plan per operator.

\begin{figure}[t]
    \centering
    \includegraphics[width=\linewidth]{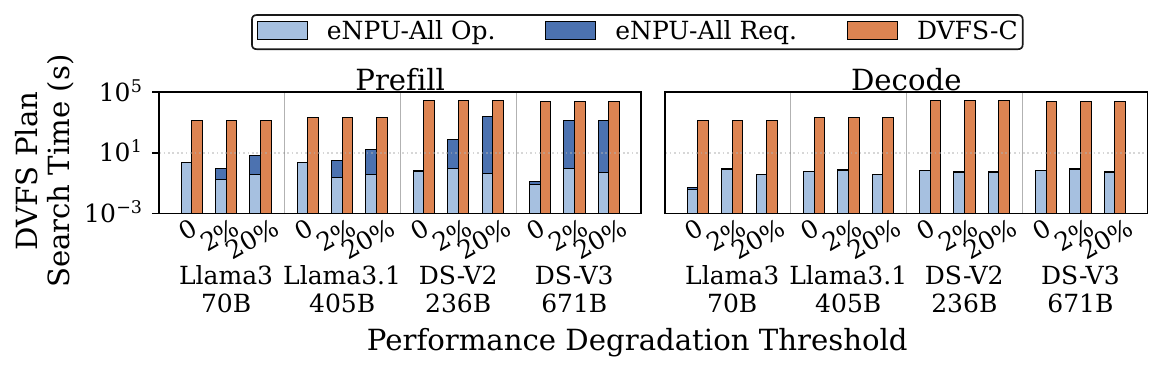}
    \caption{Compilation time overhead of \pname{}.}
    \label{fig:eval_compile_time_overhead}
\end{figure}

\noindent\textbf{Compilation time.}
\Cref{fig:eval_compile_time_overhead} evaluates the DVFS search time of compiling one LLM request with 24 Intel Xeon Gold 6136 CPU cores and 128 GB host memory. We break down the time for operator- and request-level search phases.
For each phase, the search time scales with the number of explored plans, as shown in \Cref{tab:num_explored_dvfs_plans}.
\mbox{\pname{}}'s compilation time overhead for individual operators/kernels is small (less than 10 seconds for all operators in a model); it is unlikely to bottleneck kernel development or optimization iterations.
Users can also temporarily disable \mbox{\pname{}} during development.
For each request, \pname{} takes at most 40 minutes, while \dvfsc can incur more than 8 hours of search time due to the inefficiency of genetic algorithm.
With up to 2304 pre-compiled buckets per model (see \mbox{\S\ref{sec:design:integration}}), the compilation takes 15360 core-hours ($\approx$5 hours with 3000 CPU cores, which is manageable for production LLM services deployed in datacenters).

\begin{table}[t]
\centering
  \caption{Number of explored DVFS plans with 20\% performance degradation threshold. ``Op.'' means Operator-level plans. ``Req.'' means Request-level plans.}
  \label{tab:num_explored_dvfs_plans}
  \begin{tabular}{@{}lp{0.6cm}p{0.5cm}p{0.5cm}p{0.75cm}p{0.8cm}p{0.5cm}p{0.9cm}@{}}
    \toprule
    \multirow{2}{1.1cm}{} & \multirow{2}{*}{} & \multicolumn{2}{c}{\dvfsc} & \multicolumn{2}{c}{\textit{\pname{}-All}} & \multicolumn{2}{c}{\ideal} \\
    \cmidrule(lr){3-4} \cmidrule(lr){5-6} \cmidrule(lr){7-8}
    & & Op. & Req. & Op. & Req. & Op. & Req. \\
    \midrule
    \multirow{2}{1.1cm}{Llama3 70B} 
    & Prefill &578 & 2.5M &15,345 &2881& 45M &2881\\
    & Decode  &714 & 2.5M &19,561 &1 &45M &1 \\
    \midrule
    \multirow{2}{1.1cm}{Llama3.1 405B} 
    & Prefill &578 & 2.5M &15,376 &4537 &45M &4537 \\
    & Decode  &714 &2.5M  &19,034 & 1&45M &1\\
    \midrule
    \multirow{2}{1.1cm}{DeepSeekv2 236B} 
    & Prefill &1326 & 2.5M &36,373 &118,861 &45M &118,861 \\
    & Decode  &1326 & 2.5M & 37,045&1&45M &1 \\
    \midrule
    \multirow{2}{1.1cm}{DeepSeekv3 671B} 
    & Prefill &1394 & 2.5M &37,944 & 27,451&45M &27,451 \\
    & Decode  & 1394 & 2.5M &36,890 &1& 45M&1 \\
    \bottomrule
  \end{tabular}
\end{table}

\subsection{Overhead Breakdown}
\label{sec:eval:overhead}


\noindent\textbf{Performance overhead.}
As shown in \Cref{fig:eval_request_all}, at zero performance degradation threshold, \pname{}'s performance overhead stays below 4\%.
The overhead stems primarily from V/$f$ transition delays and domain-crossing overhead.
The domain-crossing penalty would be much higher without \pname{}'s VU internal forwarding mechanism (\S\ref{sec:design:hw}); we perform a sensitivity study in \S\ref{sec:eval:vu_ifd}.
At larger thresholds, \pname{}'s performance degradation always stays below the allowed threshold,
Despite these overheads, \pname{} preserves the target SLO guarantees in end-to-end serving (see \S\ref{sec:eval:llm_service}).

\begin{figure}[t]
    \centering
    \includegraphics[width=\linewidth]{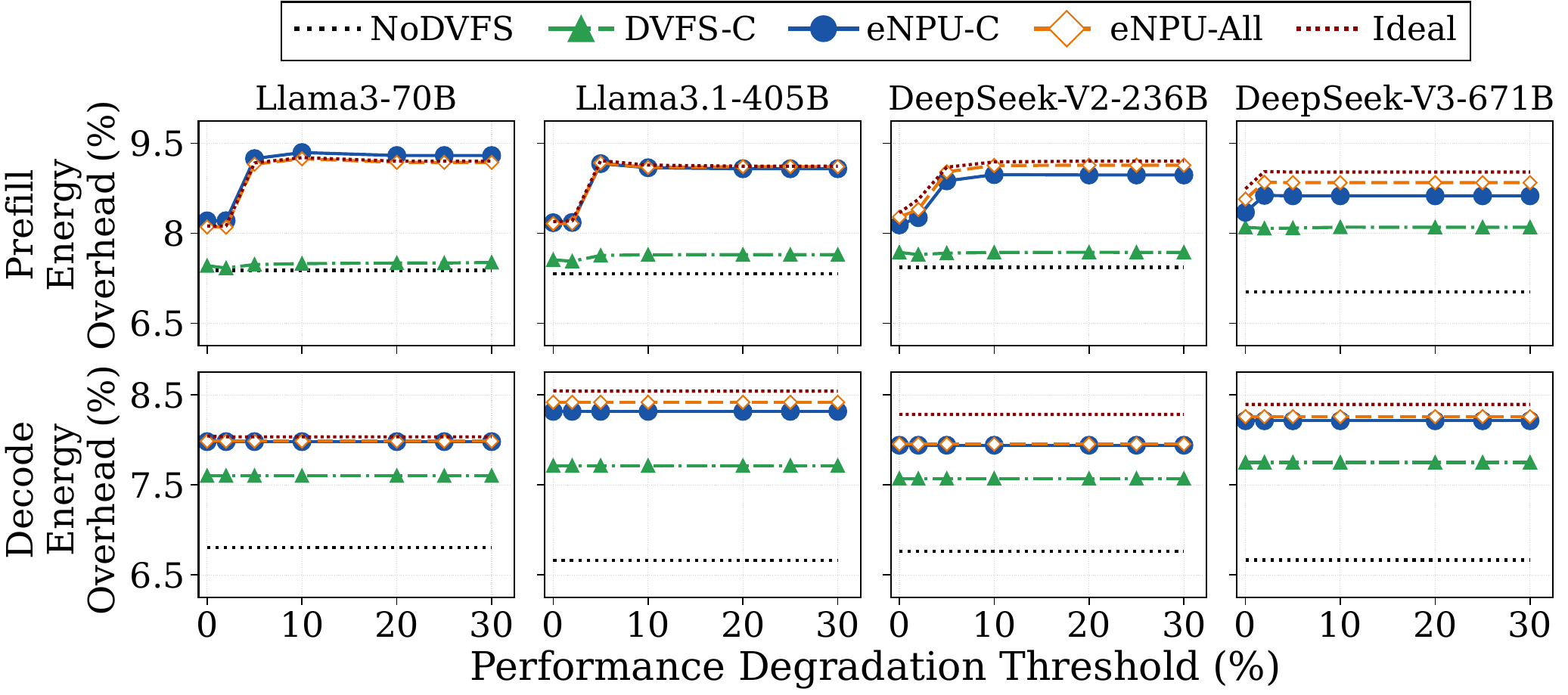}
    \caption{Energy overhead due to IVR power conversion.}
\label{fig:eval_conversion_overhead}
\end{figure}

\noindent\textbf{IVR power conversion overhead.}
\Cref{fig:eval_conversion_overhead} quantifies the energy lost due to IVR power conversion 
inefficiency.
IVR power conversion efficiency degrades at lower output voltage, so conversion overhead grows slightly as \pname{} throttles V/$f$ more aggressively at larger SLO slacks.
However, \pname{}'s net energy savings outweigh this overhead.


\subsection{Benefits of VU Shadow Vector Register}
\label{sec:eval:vu_ifd}

\begin{figure}[t]
    \centering
    \includegraphics[width=\linewidth]{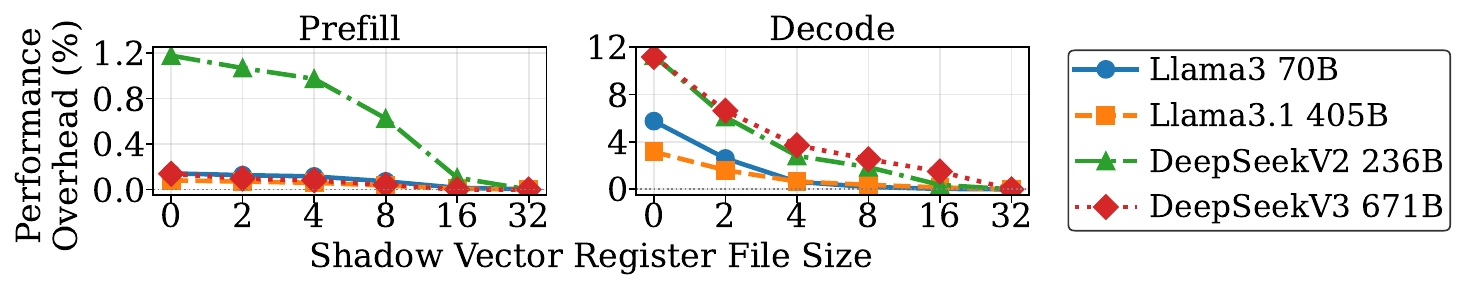}
    \caption{Impact of varying VU shadow vector register file size at performance degradation threshold 0\%.}
    \label{fig:eval_vu_ifd}
\end{figure}

\Cref{fig:eval_vu_ifd} shows how the VU shadow vector register file size (\S\ref{sec:design:hw}) impacts performance.
We focus on 0\% performance degradation threshold, as the overhead is the most obvious. For larger thresholds, the component frequencies are lower, so the overhead becomes less obvious.

A shadow regfile size of 4 achieves $<$4\% performance penalty.
Decode has a higher sensitivity to the shadow register file size, as many operators are VU-intensive.
A larger shadow regfile maps more simultaneously live VU dependencies to shadow registers, reducing frontend stalls when the dependency distance is not long enough to hide frontend-VU round-trip latency, despite its larger area and power cost.
The performance penalty is near-0 if the shadow regfile is as large as the frontend regfile (32 registers).
We empirically pick a default shadow regfile size of 4, which provides 
a reasonable trade-off between area overhead and performance.

\subsection{Sensitivity Analysis}
\label{sec:eval:sens}

\begin{figure}[t]
    \centering
    \includegraphics[width=\linewidth]{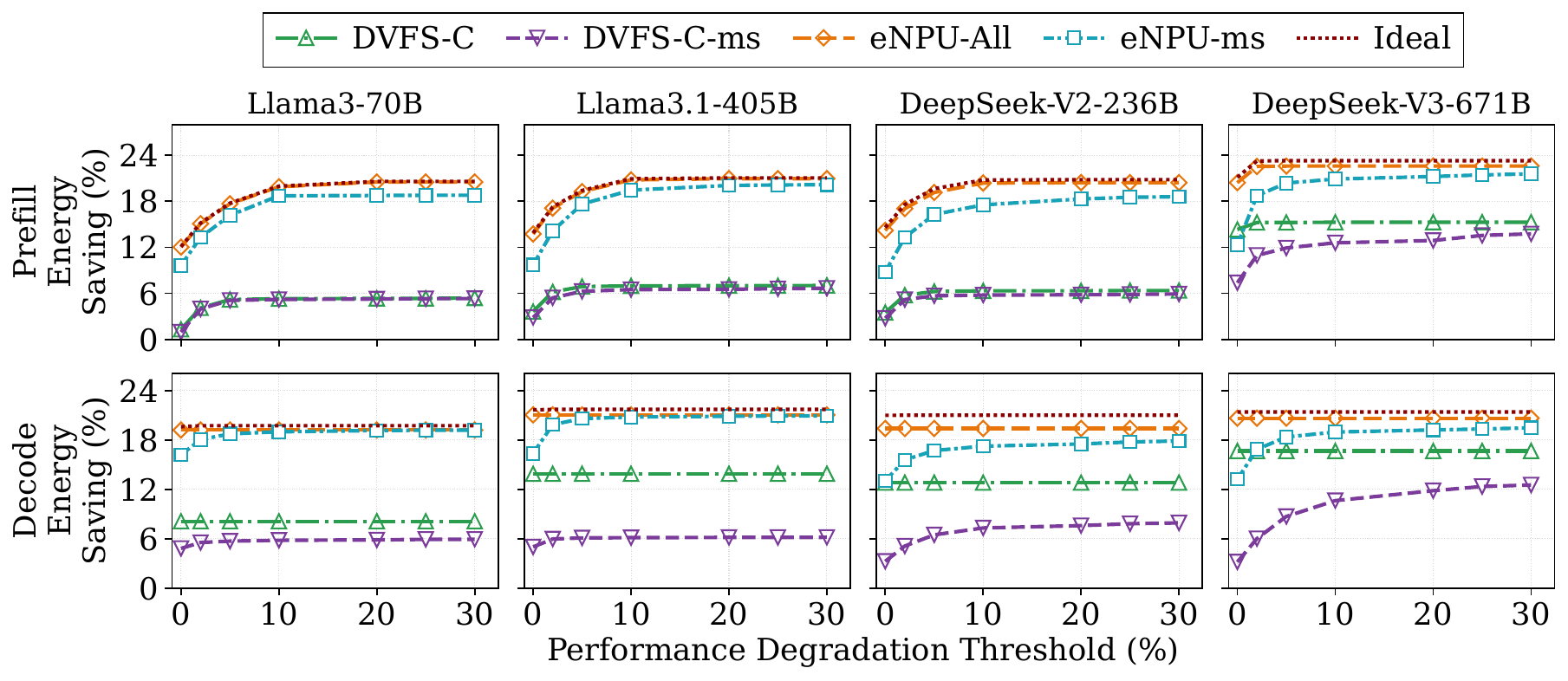}
    \caption{Energy saving of different temporal and spatial DVFS granularities.}
    \label{fig:eval_sens_dvfs_granularity}
\end{figure}

\noindent
\textbf{DVFS granularity.}
\mbox{\Cref{fig:eval_sens_dvfs_granularity}} isolates the impact of spatially and temporally fine-grained DVFS.
In \mbox{\textit{DVFS-C-ms}} and \mbox{\textit{\pname{}-ms}}, we merge consecutive sub-millisecond operators into epochs ($\geq$5 milliseconds) following the preprocessing step in the original paper~\cite{ascend_dvfs:asplos25}, and then apply DVFS-C's genetic algorithm or \mbox{\pname{}}'s DVFS policy for each epoch. \mbox{\dvfsc{}} and \mbox{\full{}} use operator-level (sub-$\mu$s-scale) DVFS.

With temporally fine-grained DVFS, {\full{}} achieves up to 8.1\% higher energy saving than {\textit{\pname{}-ms}}.
{\textit{\pname{}-ms}} applies a unified V/$f$ setting across an entire time epoch, forcing less critical operators to run at high V/$f$, while {\full{}} allows varying V/$f$ for different operators.
Even with coarse temporal granularity, {\textit{\pname{}-ms}} still saves 4.95\%--14.76\% more energy than {\textit{DVFS-C-ms}} by exploiting component-level DVFS.

\begin{figure}[t]
    \centering
    \includegraphics[width=\linewidth]{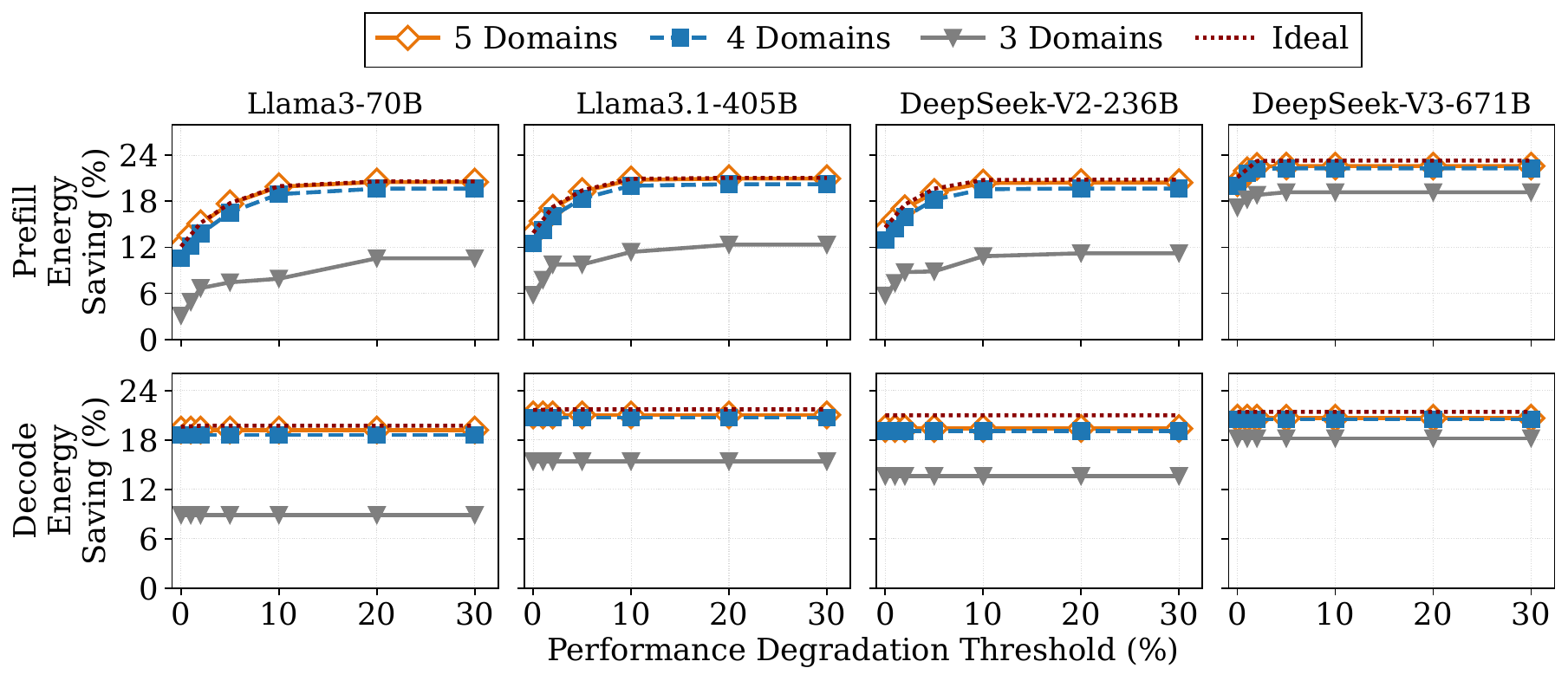}
    \caption{Energy saving of \pname{} with different V/$f$ domain configurations.}
    \label{fig:eval_sens_domain_count}
\end{figure}

\begin{figure}[t]
    \centering
    \includegraphics[width=\linewidth]{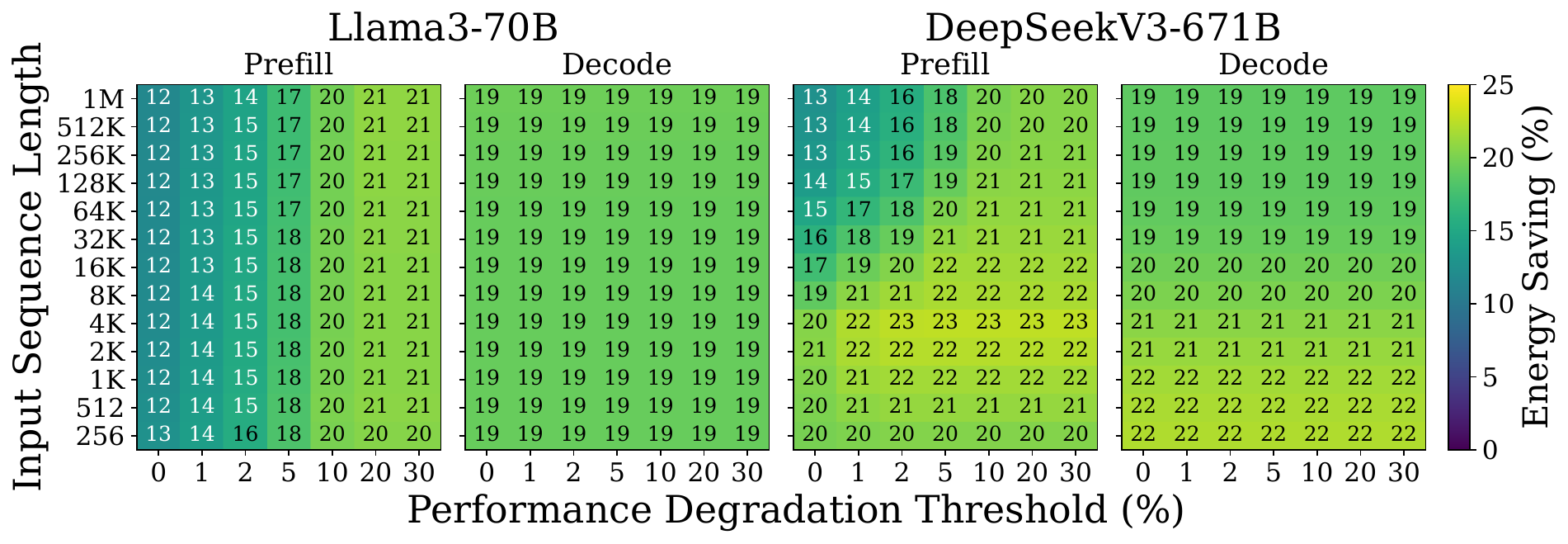}
    \caption{Energy savings of \mbox{\pname{}} over \mbox{\nodvfs{}} with various sequence lengths. We show two models due to space limits.}
    \label{fig:eval_var_seqlen}
\end{figure}

\begin{figure}[t]
    \centering
    \includegraphics[width=\linewidth]{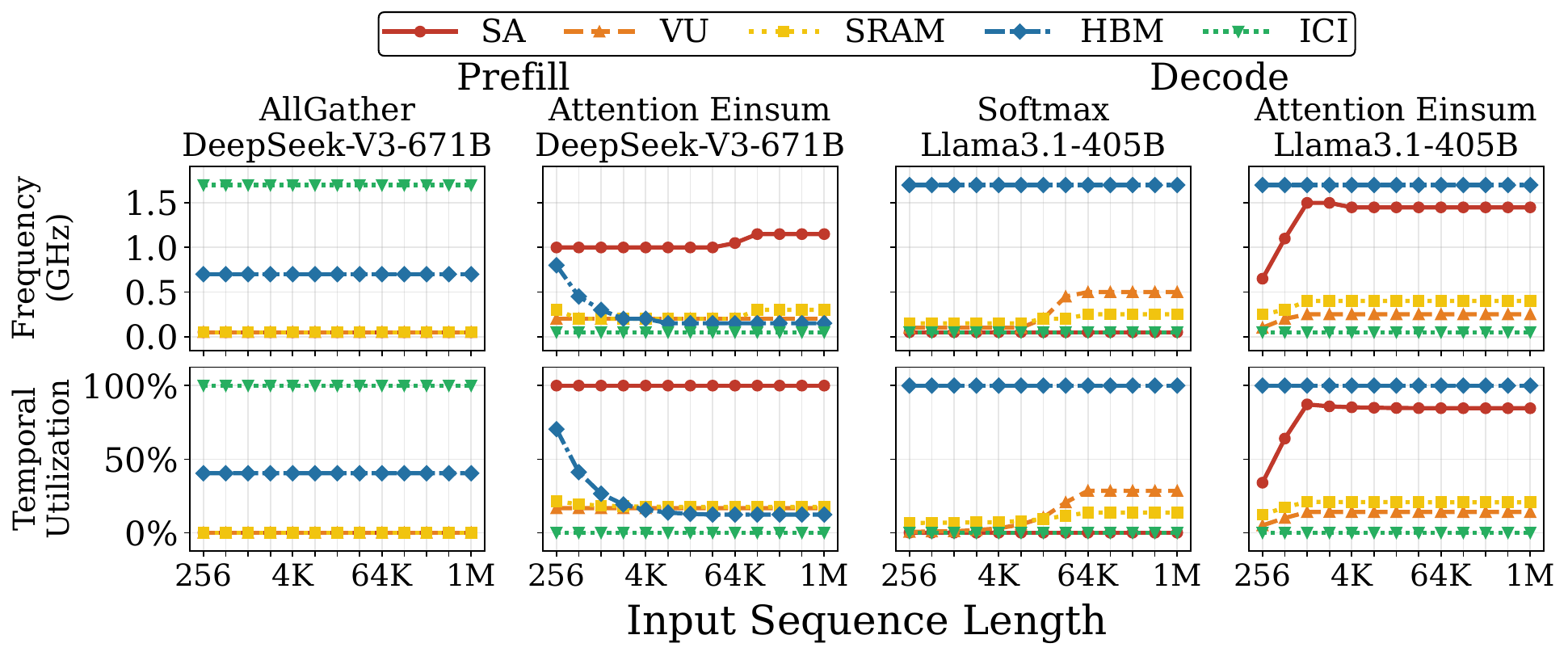}
    \caption{Frequencies selected by \mbox{\full{}} (top row) and the original component utilization without DVFS (bottom row) for representative operators, w.r.t. various sequence lengths. We plot the results for performance degradation threshold 30\%; other thresholds have a similar trend.}
    \label{fig:eval_var_seqlen_freq_util}
\end{figure}

\noindent
\textbf{V/$\bm f$ domain count.}
\mbox{\Cref{fig:eval_sens_domain_count}} studies \mbox{\pname{}}'s sensitivity to the number of V/$f$ domains. The 5-domain configuration is \mbox{\full{}}. The 4-domain version merges the frontend (FE), SA, and VU into one ``compute'' domain, while keeping the SRAM, HBM, and ICI separate. The 3-domain version forms an ``NPU core'' domain for FE, SA, VU, and SRAM, plus separate HBM and ICI ``uncore'' domains.

4-domain saves 2.3\%--11.0\% more energy than 3-domain, and 5-domain saves 0.1\%--1.5\% more than 4-domain.
The major benefit comes from separating SAs and SRAM into different V/$f$ domains, because in our simulated NPU chip (TPUv5p), SA dominates the on-chip power for prefill, and SRAM dominates that for decode (see \mbox{\Cref{fig:eval_energy_breakdown_component}}).
Hence, the energy saving from separately throttling VUs is relatively modest.
Nevertheless, the extra hardware cost of having 5 domains over 4 domains is small (only 0.64\% for TPUv5p, including the VU async FIFO, the voltage level shifters, and the shadow vector register file in \mbox{\Cref{tab:hardware_area_scale}}).

As future NPU chips incorporate more powerful VUs~\mbox{\cite{googlecloud2026tpu8t}} for increasingly VU-intensive LLM operators (e.g., Gated DeltaNet~\mbox{\cite{yang2025gateddelta,qwen2025qwen3next}}
and compressed/sparse-attentions such as HCA and CSA~\mbox{\cite{deepseekai2026deepseekv4}}), the VUs will account for a larger portion of NPU core power relative to SAs and SRAM. Similarly, for NPUs with smaller SAs (e.g., the Coral NPU core in \mbox{\Cref{tab:hardware_area_power}}), VU power is relatively larger.
In these cases, a separate VU domain would bring more energy savings.

\noindent
\textbf{Variable sequence length.}
\mbox{\Cref{fig:eval_var_seqlen}} shows the energy savings with various input sequence lengths. We keep the output sequence length at 512 tokens, as the decode performance and energy consumption depend on the total context length, which is covered by sweeping the input lengths.

The energy saving depends on the request's bottlenecking component, which depends on the model architecture and the sequence length.
For Llama3-70B prefill, the energy saving is similar across sequence lengths, as the request is compute-intensive and SA-bound.
For DeepSeekV3-671B prefill, when we increase sequence length from 256 to 4K, the request is ICI-bound; it cannot saturate the 128$\times$128 SA, but the ICI overhead grows near-linearly. Hence, the SA temporal utilization (similarly for VU and SRAM) drops relatively, exposing more throttling opportunities.
As we increase sequence length beyond 4K, the SA, VU, and SRAM busy time increase quadratically due to the attention layer.
Hence, the energy savings from SAs, VUs, and SRAM shrink, leading to relatively smaller total energy savings.
The takeaway is similar for decode: Llama3-70B is always memory-bound, so the energy saving is similar across sequence lengths; DeepSeekV3-671B is ICI-bound for shorter sequence lengths ($\leq$ 4K) and becomes more memory-bound for longer sequences, so the energy saving drops slightly due to the near-linear power-frequency relationship of HBM (see \Cref{fig:synth_peak_power}).

In \mbox{\Cref{fig:eval_var_seqlen_freq_util}}, we investigate how sequence lengths affect the DVFS plans generated by \mbox{\pname{}}. In general, the selected frequency of a component is proportional to its temporal utilization measured at peak frequency (i.e., \mbox{\nodvfs{}}).
This is expected, as a highly utilized component has less slack in the operator schedule and hence must run at a higher frequency to meet the performance target. In contrast, a less-utilized component is often off the critical path, so it can be throttled with little impact on end-to-end latency.





\begin{figure}[t]
    \centering
    \includegraphics[width=\linewidth]{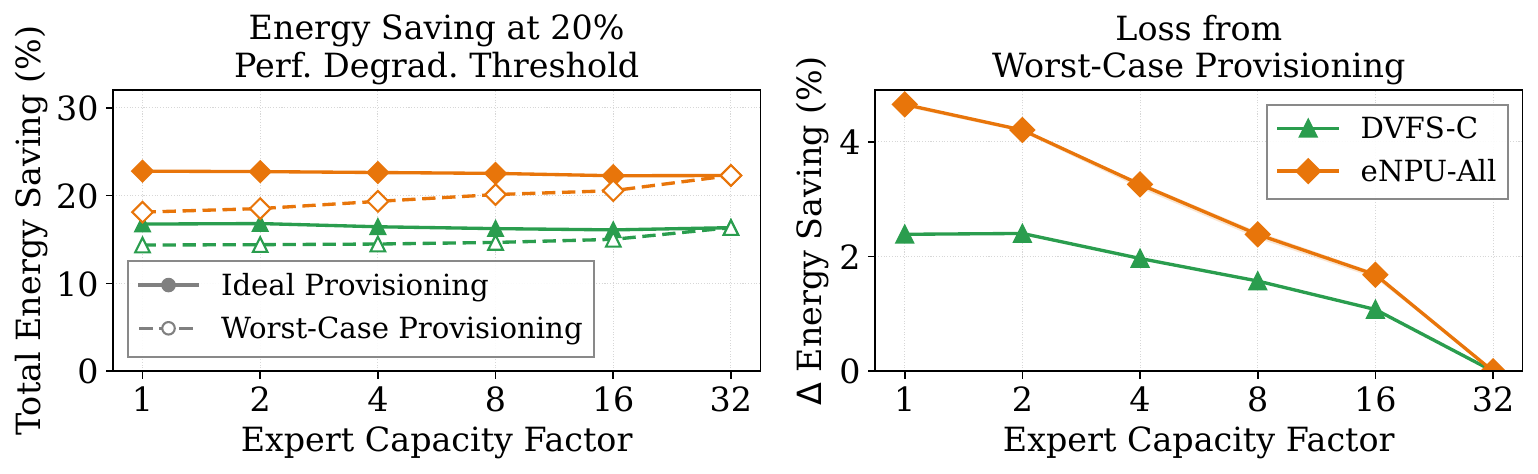}
    \caption{Left: Energy savings over \mbox{\nodvfs{}} at different expert load imbalances. Right: Energy saving difference between ideal and worst-case provisioning, averaged across 0\%--20\% performance degradation thresholds (this threshold has negligible impact on the energy saving difference). We only show DeepSeekV3-671B prefill due to space limits. The energy saving in decode is not affected by load imbalances since the number of tokens is very small.}
    \label{fig:eval_sens_moe}
\end{figure}

\noindent
\textbf{MoE expert load imbalance.}
By default, \mbox{\pname{}} selects DVFS plans based on the worst-case expert load imbalance (\mbox{\S\ref{sec:design:integration}}), which ``overprovisions'' energy budget if the real token distribution is more balanced. In \mbox{\Cref{fig:eval_sens_moe}}, we study different token distributions by varying the simulated expert capacity factor, and compare \mbox{\pname{}} with an ideal provisioning that perfectly predicts the load imbalance and incurs no SLO violations.
When the load is perfectly balanced (expert capacity factor $=1$), \mbox{\pname{}} will save 4.7\% less energy, but it still achieves 18\% energy saving over \mbox{\nodvfs{}}. Regardless of the load imbalance, \mbox{\pname{}} always saves more energy than \mbox{\dvfsc{}}.

\subsection{Benefits of DVFS $+$ Power Gating}
\label{sec:eval:dvfs_pg}

\begin{figure}[t]
    \centering
    \includegraphics[width=\linewidth]{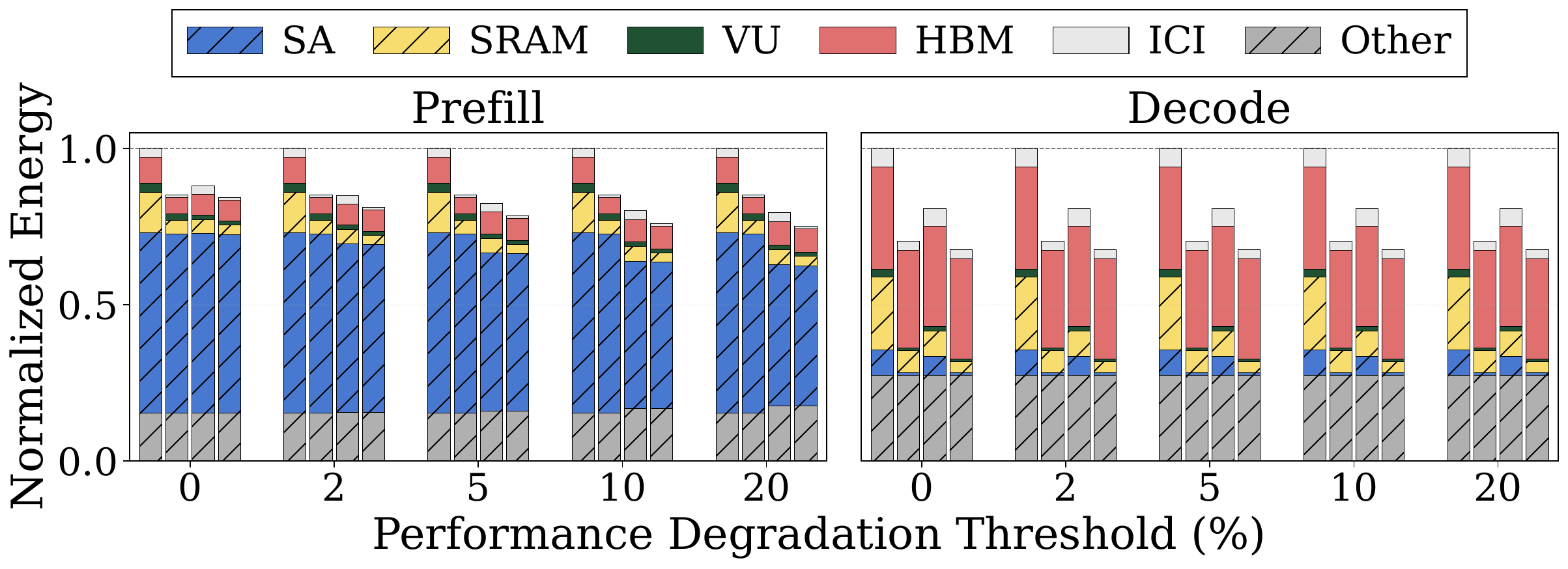}
    \caption{Energy savings with DVFS and power gating (PG). The bars from left to right are None (no DVFS or PG), PG-Only, DVFS-Only, and PG+DVFS.}
    \label{fig:eval_energy_pg}
\end{figure}



\Cref{fig:eval_energy_pg} quantifies the benefits of combining DVFS and power gating (PG).
We compare None (no DVFS or PG), PG-Only, DVFS-Only, and PG+DVFS.
We apply the state-of-the-art NPU PG solution ReGate~\cite{regate:micro25}, which uses the compiler to analyze component idleness and orchestrate PG at instruction level.
We first apply \pname{}'s DVFS algorithm to each operator, and then apply ReGate to instrument power gating commands based on the VLIW code sequence after DVFS.
When the performance degradation threshold is 0, PG-Only outperforms DVFS-only.
This is because PG-Only can exploit component underutilization by throttling the supply voltage to near-zero, yielding more static power reductions than DVFS-Only (which only throttles voltage to 0.45V).
However, the unique advantage of DVFS lies in its ability to actively trade excess performance for energy savings.
At large SLO slacks for prefill, DVFS-Only outperforms PG-Only.
By combining DVFS and PG, we can achieve maximum energy savings (up to 24.9\%/32.4\% for prefill/decode).

\subsection{Energy Savings for LLM Service}
\label{sec:eval:llm_service}

\begin{figure}[t]
    \centering
    \includegraphics[width=\linewidth]{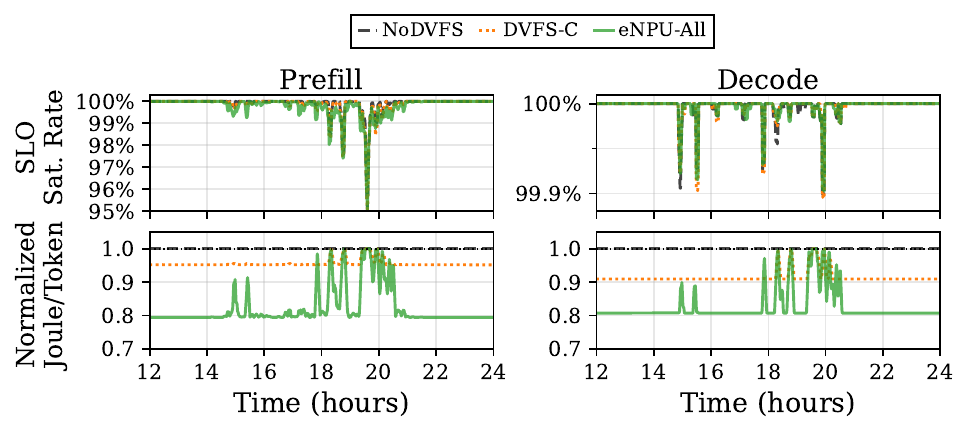}
    \caption{End-to-end energy savings for LLM serving. We show 12--24 hours, as the trend is flat during off-peak hours.}
    \label{fig:eval_energy_workload}
\end{figure}

In \Cref{fig:eval_energy_workload}, we evaluate the end-to-end energy savings of \pname{} for a 24-hour production LLM serving trace.
The trace includes both peak hours (14--20 hours) and off-peak hours.
We employ \full and \dvfsc using the same DVFS-aware request scheduling mechanism and safeguarding rules (\S\ref{sec:design:integration}).
We study Llama3-70B as an example. The insights apply to other models as well.

\full only needs $\approx$10\% SLO slack to achieve maximum energy savings.
During off-peak hours, \dvfsc and \full achieve 4.8\%/9.0\% and 20.5\%/19.3\% energy savings for prefill/decode with no obvious drop in SLO satisfaction rate.
During peak hours, the available SLO slack shrinks, so DVFS is applied less aggressively.
In \mbox{\full{}}, 12\%/10\% of prefill/decode batches during peak hours use a less aggressive DVFS plan (safeguard rule 1 in \mbox{\S\ref{sec:design:integration}}), and only 2.3\%/1.1\% of prefill/decode batches are locked to the peak clock speed (rule 2).
Nevertheless, \full achieves much higher energy savings than \dvfsc under limited SLO slacks (16.4\%/16.3\% vs. 4.2\%/8.2\% for prefill/decode).




%% file: discussion.tex
\section{Discussion}
\label{sec:discussion}

\noindent
\textbf{Adapting \mbox{\pname{}} for overclocking.}
\mbox{\pname{}}'s DVFS policy (\mbox{\S\ref{sec:design:sw_dvfs}}) can be adapted to overclock an NPU chip given a power budget.
The operator-level exploration can cover overclocking options for each component.
Then, the request-level search can work the same way, starting from the fastest plan and iteratively throttling each component's frequency until the DVFS plan satisfies the power budget.

\noindent
\textbf{Implications of advanced technology nodes.}
In our prototype of \mbox{\pname{}}, we use ASAP7 PDK as a reproducible case study, which roughly matches the technology nodes of TPUv4/TPUv5p. As long as the $P \propto V^2f$ law continues to hold, DVFS will remain an effective technique for future nodes.
Moreover, as different components scale at different paces (e.g., logic scales faster than SRAM) and have different performance requirements, their V/$f$ and power characteristics will remain diverse. \pname{} offers a solution to manage such component diversity and maximize DVFS opportunities.

\noindent
\textbf{Prefill/decode disaggregation vs. mixed continuous batching.}
In our evaluation, we employ prefill/decode (P/D) disaggregation following state-of-the-art LLM serving systems~\mbox{\cite{splitwise:isca24,DistServe:osdi24}}, as these two phases have distinct resource demands.
\mbox{\pname{}} can also be integrated with aggregated P/D serving (i.e., mixed continuous batching~\mbox{\cite{sarathiserve:osdi24}}) in the same way as described in \mbox{\S\ref{sec:design:integration}}, as it picks the DVFS plan for each batch, no matter whether it is a batch of prefill requests or decode requests.
The per-batch energy saving trend would be similar to P/D disaggregation, as prefill batches are typically still compute-bound and decode batches are memory-bound.


%% file: related_work.tex
\section{Related Work}
\label{sec:related}

\noindent
\textbf{DVFS techniques.} 
Prior work explored different DVFS granularities and policies.
They often treat an entire chip as one V/$f$ domain, or partition V/$f$ domains only across replicated, homogeneous units such as CPU cores~\cite{dvfs_per_core:hpca08} or GPU SM, and core/uncore domains~\cite{dvfs_uncore_haswell:ipdpsw15,gpu_wattch:isca13,dvfs_gpu_mem:hpca18}.
In contrast, \pname{} enables independent DVFS across heterogeneous components within an NPU core.
Prior work explores dynamic schemes based on runtime feedback~\cite{dvfs_pid:asplos04,dvfs_mcd:hpca02,dvfs_phase_prediction:micro06,dvfs_compiler_dynamic:micro05,dvfs_dynamic:tcad2005,dvfs_orig:osdi94} and static schemes based on offline profiling or compiler analysis~\cite{dvfs_profile:pldi03,dvfs_compiler:pldi03,dvfs_static_ilp:lctes02}.
They assume one coarse-grained V/$f$ domain or a few homogeneous domains, which cannot directly apply when we must jointly optimize V/$f$ choices for multiple components and account for their interaction with VLIW instruction scheduling.
\pname{} addresses this challenge with a compiler-driven greedy search.

\noindent
\textbf{Power management techniques for AI chips.}
Recent works have explored both static and dynamic power management for AI chips. 
PCSTALL~\cite{dvfs_gpu:amd:asplos23} uses a program counter-based prediction mechanism to configure DVFS on GPUs. 
The Ascend NPU~\cite{ascend_dvfs:asplos25} employs a genetic algorithm to search for the optimal AI core frequency.
\pname{} enables component-level V/$f$ domains on NPUs and addresses the unique HW and SW challenges of exploiting the new DVFS optimization space.
UPTPU~\mbox{\cite{uptpu}} power gates MAC units in spatially underutilized systolic arrays.
ReGate~\cite{regate:micro25} enables compiler-directed power gating across all major NPU components.
\pname{} is complementary to these power gating techniques.


\noindent
\textbf{Power/energy-aware LLM serving.}
Power/energy efficiency of LLM serving has become a critical concern.
Proposed optimizations include request routing techniques that account for renewable energy~\cite{heron:arxiv25} or power hotspots across server racks~\cite{tapas:asplos25}.
Many works scale down GPU frequency to exploit SLO slack and save energy, using offline profiling~\cite{dynamollm:hpca25,throttll:hpca25}, online feedback-based control~\cite{voltanallm:arxiv25,userve:atc24}, and ILP solvers~\cite{heron:arxiv25} to find the optimal frequency.
They are designed for whole-chip DVFS on GPUs and cannot directly apply to spatially fine-grained DVFS on NPUs.
Prior studies also examine model placement strategies, such as varying the tensor parallelism degree~\cite{tensorparallel:arxiv19} to enable latency-energy trade-offs~\cite{userve:atc24,dynamollm:hpca25,throttll:hpca25,biscale:arxiv26,ecoserve:arxiv25} or grouping operators by their sensitivity to chip frequency~\cite{userve:atc24}.
Compared to these system-level optimizations, \pname{} addresses the unique HW and SW challenges and provides a holistic solution for exploiting component-level DVFS on NPU chips for LLM serving.

%% file: conclusion.tex
\section{Conclusion}
\label{sec:conclusion}

We identify the DVFS opportunities for LLM serving on NPUs and propose \pname{}, which enables component-level DVFS on NPUs. \pname{} improves the energy efficiency of LLM serving while preserving strict SLO guarantees.

%% file: ack.tex
\section*{Acknowledgements}

We thank the anonymous reviewers for their valuable feedback.
We thank Shaobo Li, Eric Zhou, Yikang Yue, Jeff Zhang, and Anna Li from the Systems Platform and Intelligence Lab (Illinois PlatformX) for proofreading the paper and improving the presentation of the figures.
This work was partially supported by Google’s TPU Research Cloud (TRC) program and NSF grants CAREER CNS-2144796 and CCF-2107470.

%% file: ae.tex
\clearpage
\appendix
\section{Artifact Appendix}

\subsection{Abstract}

This artifact contains the NPU simulator used to produce the major motivation study (\S\ref{sec:study}) and evaluation results (\S\ref{sec:eval}) in our paper.
The simulator is based on the open-source NeuSim codebase (\url{https://github.com/platformxlab/NeuSim}).
Our artifact provides automated scripts, input traces, and necessary instructions
to reproduce these figures. The artifact can be executed on any x86 machine. 

\subsection{Artifact check-list (meta-information)}


{\small
\begin{itemize}
  \item {\bf Data set: } \url{https://github.com/Azure/AzurePublicDataset}
  \item {\bf Run-time environment:} Ubuntu 24.04, Python 3.12.
  \item {\bf Hardware:} Any x86 machine. We recommend at least 24 CPU cores and 64 GB memory.
  \item {\bf Output:} Simulator output statistics and plots.
  \item {\bf How much disk space required (approximately)?:} 20 GB.
  \item {\bf How much time is needed to prepare workflow (approximately)?:} $\leq$10 minutes.
  \item {\bf How much time is needed to complete experiments (approximately)?: } 12 hours with 24 CPU cores and 64 GB memory.
  \item {\bf Publicly available?: } \url{https://github.com/platformxlab/NeuSim/tree/micro26ae_enpu}
  \item {\bf Code licenses (if publicly available)?: } Apache 2.0.
  \item {\bf Archived (provide DOI)?: } \url{https://doi.org/10.5281/zenodo.21542470}
\end{itemize}
}

\subsection{Description}

\subsubsection{How to access}
The source code and traces can be downloaded from
Zenodo at \url{https://doi.org/10.5281/zenodo.21542470}. For the latest
version, we recommend visiting the \texttt{micro26ae\_enpu} branch of the NeuSim GitHub repository: 
\begin{center}
    \small\url{https://github.com/platformxlab/NeuSim/tree/micro26ae_enpu}.
\end{center}


\subsubsection{Hardware dependencies}
The artifact can be executed on any x86 machine.
We recommend at least 24 physical CPU cores, 32 GiB of
memory (64 GiB preferred), and 20 GiB of free disk space.

\subsubsection{Software dependencies}
The artifact needs a Linux environment (preferably Ubuntu 24.04 or newer), uses \texttt{pip} to manage software dependencies, and requires Python $\geq$3.12.

\subsubsection{Data sets}
The artifact uses a truncated 1-day trace from the AzurePublicDataset~\cite{dynamollm:hpca25}. The truncated trace is supplied with this artifact.

\subsection{Installation}

To retrieve the artifact, clone the repository and switch to the \texttt{micro26ae\_enpu} branch:
\begin{lstlisting}[style=BashStyle,escapechar=~~]
    git clone https://github.com/platformxlab/NeuSim.git
    cd NeuSim
    git checkout micro26ae_enpu
\end{lstlisting}

We strongly recommend following the \texttt{micro26ae.md} for detailed instructions on installing the artifact and running experiments. We list the key steps below. 
Run all commands from the NeuSim repository root:
\begin{enumerate}
    \item Install the NeuSim package and dependencies:
    \begin{lstlisting}[style=BashStyle,escapechar=~~]
        conda create --name neusim-dvfs-ae python=3.12.2
        conda activate neusim-dvfs-ae
        python -m pip install -e .
        export PYTHON=python
    \end{lstlisting}

    \item Retrieve the supplementary DVFS lookup cache and the input trace file:
    \begin{lstlisting}[style=BashStyle,escapechar=~]
SUPPLEMENTARY_DIR=/tmp/micro26ae_supplementary_files
git clone --depth 1 \
  https://github.com/XZman/micro26ae_supplementary_files.git \
  "$SUPPLEMENTARY_DIR"
(cd "$SUPPLEMENTARY_DIR" && sha256sum -c SHA256SUMS)

unzip -q -o \
  "$SUPPLEMENTARY_DIR/AzureLLMInferenceTrace_code_1day.zip" \
  -d artifact_eval/micro26/data/azure
unzip -q -o \
  "$SUPPLEMENTARY_DIR/dvfs_lookup_dvfsc.zip" \
  -d artifact_eval/micro26/data
unzip -q -o \
  "$SUPPLEMENTARY_DIR/dvfs_lookup_enpu_all.zip" \
  -d artifact_eval/micro26/data
    \end{lstlisting}

    \item Test the installation and verify hash of supplementary files:
    \begin{lstlisting}[style=BashStyle,escapechar=~]
        python -c 'import neusim; print(neusim.__version__)'
        ./artifact_eval/micro26/run_all.sh --list
        sha256sum -c artifact_eval/micro26/data/SHA256SUMS
    \end{lstlisting}
    These three commands should report NeuSim version 0.1.0, list 11 ready
    chip-level figures, and validate all three extracted inputs. See
    \texttt{micro26ae.md} for an optional 1--3 minute smoke test.
\end{enumerate}

\subsection{Experiment workflow}

Please see \texttt{micro26ae.md} for detailed instructions of the experiment workflow.
We summarize the key steps below.

\noindent
\textbf{Request-level experiments.}
The following command reproduces Figures 2, 3, 4, 11--13, 16--18, 20, and 21.
It takes approximately 6--8 hours on a 24-core machine.

\begin{lstlisting}[style=BashStyle,escapechar=~]
    CHIP_OUT=artifact_eval/micro26/reproduced/chip
    ./artifact_eval/micro26/run_all.sh \
      --results-dir "$CHIP_OUT" \
      --output-dir "$CHIP_OUT" \
      --jobs 24 --trace-workers 4 \
      --group-trace-workers standard_sweep=8 \
      --group-trace-workers domain_count=8 \
      --group-trace-workers temporal_granularity=8 \
      --group-trace-workers fixed_sequence_sweep=24 \
      --group-trace-workers expert_imbalance=1 \
      --group-trace-workers power_gating=2 \
      --allow-current-ideal --resume
\end{lstlisting}

\noindent
\textbf{LLM service-level experiments.}
The following command reproduces Figures 5 and 22 by replaying the supplied
one-day trace. It takes approximately 4 hours.

\begin{lstlisting}[style=BashStyle,escapechar=~]
FLEET_OUT=artifact_eval/micro26/reproduced/fleet
./artifact_eval/micro26/run_fleet.sh \
  --output-dir "$FLEET_OUT" \
  --resume
\end{lstlisting}

The \texttt{--resume} option validates and reuses completed stages. Detailed
commands, including lookup-cache regeneration, are provided in
\texttt{micro26ae.md}.

\subsection{Evaluation and expected results}

Successful execution produces the following review reports and figures:

\begin{itemize}
  \item Request-level report:
    \path{artifact_eval/micro26/reproduced/chip/FIGURE_REVIEW.md}.
  \item Request-level PDFs:
    \path{artifact_eval/micro26/reproduced/chip/figures/}.
  \item LLM service-level report:
    \path{artifact_eval/micro26/reproduced/fleet/FIGURE_REVIEW.md}.
  \item LLM service-level PDF (Figure 5):
    \path{artifact_eval/micro26/reproduced/fleet/figure05/figures/figure_05_slo_slack.pdf}.
  \item LLM service-level PDF (Figure 22):
    \path{artifact_eval/micro26/reproduced/fleet/figure22/figures/figure_22_fleetsim_dvfs_timeseries.pdf}.
\end{itemize}

\subsection{Experiment customization}

The request-level (chip-level) simulation supports selecting figure subsets; changing models,
phases, sequence lengths, NPU chip counts and DP/TP/PP/EP configurations;
adjusting performance-degradation thresholds, MoE expert imbalance, DVFS
domain/temporal granularity, and power gating; and modifying hardware V/$f$ and
power curves.

The LLM service-level (fleet-level) simulation supports custom CSV traces,
custom SLO targets, and custom prefill/decode fleet sizes and NPU instance configurations.
These require re-generating a fresh DVFS lookup cache.
Please see more details in \texttt{micro26ae.md}.


\subsection{Methodology}

Submission, reviewing and badging methodology:

\begin{itemize}
  \item \url{https://www.acm.org/publications/policies/artifact-review-and-badging-current}
  \item \url{https://cTuning.org/ae}
\end{itemize}

